\documentclass[a4paper,fleqn,usenatbib]{mnras}

\usepackage{newtxtext,newtxmath}

\usepackage[T1]{fontenc}
\usepackage{ae,aecompl}

\usepackage{graphicx}	
\usepackage{amsmath}	
\usepackage{amssymb}	

\usepackage{mathrsfs}
\usepackage{subfigure}

\usepackage[utf8]{inputenc}
\usepackage[authoryear]{natbib}
\usepackage{multicol}

\title[A parallel-GPU code for asteroid aggregation]{A parallel-GPU code for asteroid aggregation problems with angular particles}

\author[F. Ferrari et al.]{
	Fabio Ferrari,$^{1,2}$\thanks{E-mail: fabio1.ferrari@polimi.it / fabio.ferrari@jpl.nasa.gov}
	Mich\`ele Lavagna,$^{1}$
	and Emmanuel Blazquez$^{1}$
	\\
	$^{1}$Department of Aerospace Science and Technology, Politecnico di Milano, Via La Masa 34, Milan, 20156, Italy\\
	$^{2}$Jet Propulsion Laboratory, California Institute of Technology, 4800 Oak Grove Dr, Pasadena, 91109, US\\
}

\date{Accepted XXX. Received YYY; in original form ZZZ}

\pubyear{2019}

\begin{document}
\label{firstpage}
\pagerange{\pageref{firstpage}--\pageref{lastpage}}
\maketitle

\begin{abstract}
The paper presents a numerical implementation of the gravitational N-body problem with contact interactions between non-spherically shaped bodies. The work builds up on a previous implementation of the code and extends its capabilities. The number of bodies handled is significantly increased through the use of a CUDA/GPU-parallel octree structure. The implementation of the code is discussed and its performance are compared against direct N$^2$ integration. The code features both smooth (force-based) and non-smooth (impulse-based) methods, as well as a visco-elastic non-smooth method, to handle contact interaction between bodies. The numerical problem of simulating ``rubble-pile'' asteroid gravitational aggregation processes is addressed. We discuss the features of the problem and derive criteria to set up the numerical simulation from the dynamical constraints of the combined gravitational-collisional problem. Examples of asteroid aggregation scenarios that could benefit from such implementation are finally presented.
\end{abstract}

\begin{keywords}
methods: numerical -- minor planets, asteroids: general -- gravitation -- planets and satellites: dynamical evolution and stability -- planets and satellites: formation
\end{keywords}



\section{Introduction}
	The numerical resolution of the N-body problem is one of the most popular problems in high performance computing. The intrinsic $O(N^2)$ complexity of the problem has challenged many astronomers and computer scientists since the beginning of the computing era. N-body algorithms have evolved side by side with the increase of computing power and availability of new resources, such as Graphic Processing Units (GPUs). State-of-the-art implementations include parallel N-body tree codes~\citep{Nbodysims1,Nbodysims2,Nbodysims3}, hybrid codes~\citep{Aarseth1999,Aarseth,Wang2015}, adaptive algorithms of optimal orders~\citep{Pruett}, systolic algorithms~\citep{Dorband}, or more generally symplectic codes~\citep{Wisdom,Duncan,Chambers}. In order to be able to solve granular dynamics problems, the capability of gravitational N-body codes needs to be extended to include contact interaction between non-point-like bodies. Two main classes of methods are commonly used to deal with contact interactions: hard- and soft-body methods. The first consider impulsive contacts between nondeformable particles~\citep{Alder1959,Jean1987}. This method is numerically very stable and allows to simulate effectively the dynamics of hundreds of thousands of bodies. In more recent years, the alternative class of soft-body methods have been developed~\citep{Cundall1979}. Based on a force-driven approach, soft-body methods are more suitable than hard-body ones for the simulation of smooth dynamics. These could be used to complement some deficiencies of hard-body models. For example, due to the impulsive nature of contact exchanged, hard-body methods are not adequate to reproduce phenomena such as wave propagation in granular media~\citep{Gilardi2002}. However, the numerical solution of soft-body methods is more unstable and often requires extremely small time steps to adequately reproduce elastic forces at contact. Both hard- and soft-body models have their own advantages and shortcomings and their use shall be carefully weighted upon the application scenario. As a general rule, hard-body methods qualify to simulate non-smooth dynamics, and vice versa: soft-body is suitable for smooth problems. However, whether a problem should be classified in either one category is a controversial field of debate: both models rely on sets of non-directly measurable parameters, which often offer poor physical interpretation when compared to real-world scenarios~\citep{Dubois2018}. To date, both hard- and soft-body methods have been successfully used to simulate a wide variety of planetary science scenarios: \cite{Richardson} used their hard-body model to simulate gravitational aggregation of rubble-pile asteroids and planetary ring dynamics~\citep{Porco2008}; \cite{Wada2006} used a soft-body approach to study cratering impacts under constant gravity field and more recently, soft-sphere methods with mutual gravity between bodies have been used to simulate rubble-pile scenarios~\citep{Sanchez2011,Tancredi2012,Schwartz2012}. Another key aspect for a realistic evaluation of contact interactions concerns the shape of interacting bodies. All aforementioned codes are limited to spherically-shaped bodies. The use of spherical shapes is now considered to be a very relevant simplification for problems where contact between bodies happens. The use of spheres allows to drastically reduce the computational burden, but can heavily affect the realism of the model when solving for the contact interaction between bodies~\citep{Michel5}. As such, the determination of the final shape of the aggregates and fragments could not be addressed~\citep{Richardson2}. Few attempts have been made using hard-body polyhedral grains together with video game engines to solve for contact interactions~\citep{Movshovitz2012}. Although polyhedral bodies provide a better modeling, compared to spheres, of some physical phenomena (e.g. related to interlocking between bodies) the computational accuracy provided by game engines is very low (single precision) and not able to satisfy basic energy or angular momentum conservation requirements. Recently~\cite{Fabio} implemented the N-body aggregation problem with non-spherical shapes and non-smooth contacts, using modules of Chrono::Engine (C::E)~\citep{Chrono,Tasora2016}, an open software optimized for granular and multi-body engineering problems able to handle contacts and collisions of a large number of complex-shaped bodies. The results obtained were very promising and satisfactorily accurate in terms of energy and angular momentum conservation. In their work,~\cite{Fabio} made use of direct N-body simulations to solve for gravitational dynamics and only few thousands of bodies could be handled by the simulator.
	
	The work presented here builds upon the work by~\cite{Fabio} and discusses the implementation of a parallel CUDA-GPU octree structure to extend the capabilities of code to a higher number of complex-shaped bodies. As its parent-code, the implementation is based on a re-work of the C::E multi-physics simulation engine to simulate collisions between non-spherical bodies and integrate the dynamics of the problem. The code features both non-smooth (impulse-based) and smooth (force-based) methods to handle contact interactions. Also, a compliant non-smooth method is featured and proposed for the first time for planetary science application, in the attempt to joint the advantages of numerical stability provided by its impulse-based formulation, together with its ability of simulating compliance at contact level. Section~\ref{sec:code} discusses the general architecture of the code and presents methods available to solve gravitational and collisional dynamics. The performance of the code are discussed in Section~\ref{sec:validation}, where the results of validation/test problems are reported. The applicability to asteroid aggregation problems is eventually discussed through simulation examples. We address the problem of rubble pile asteroid aggregation processes in section~\ref{sec:asteroid}. The physics of gravitational and collisional processes involved are reviewed and their numerical modeling is discussed. We derive a set of qualitative and quantitative requirements from dynamical and numerical constraints of the gravitational-collisional problem.

\section{Numerical implementation}
\label{sec:code}
	The code is built using a modular approach, taking advantage of the object-oriented C++ architecture of C::E library. Modules represent interchangeable units of software, easily accessible by the user. Each module contains methods and routines to execute specific tasks. The main modules developed and/or directly retrieved by C::E libraries are:
	\begin{multicols}{2}
		\begin{enumerate}
			\item gravity \label{enum:mod-i}
			\item contact \label{enum:mod-ii}
			\item rigid-body dynamics \label{enum:mod-iii}
			\item body creation \label{enum:mod-iv}
			\item numerical solvers \label{enum:mod-v}
			\item data input/output \label{enum:mod-vi}
			\item visual interface \label{enum:mod-vii}
			\item post processing \label{enum:mod-viii}
		\end{enumerate}
	\end{multicols}
	Modules~\ref{enum:mod-ii},~\ref{enum:mod-v},~\ref{enum:mod-vii} are directly adapted from C::E libraries, modules~\ref{enum:mod-iii},~\ref{enum:mod-iv} are based on C::E libraries and further extended with additional methods and routines, modules~\ref{enum:mod-i},~\ref{enum:mod-vi},~\ref{enum:mod-viii} are fully developed by the authors. Modules~\ref{enum:mod-vi} and~\ref{enum:mod-vii} provide routines that allow the user to interface with the code at a higher level than software implementation level. The visual interface is based on Irrlicht library~\citep{Irrlicht} and can be set to highlight desired geometrical or simulation features (forces, contacts), as well as to display real-time simulation outputs. The data input/output interface is straightforward: every user-tunable parameter of the model can be assigned via an input text file that is parsed and acquired by the code, while output data is typically saved into dedicated text files. This section presents a general overview of the numerical methods available in each module, focusing on those relevant to the gravitational aggregation problem.
	
	\subsection{Gravitational dynamics}
	\label{sub:gravity}
		Module~\ref{enum:mod-i} encloses methods and routines used to compute the gravity forces acting on each body. These are based on the evaluation of gravity field produced by point-mass or shape-based (e.g. polyhedron, ellipsoid) gravity sources. To simulate the problem of gravitational aggregation, the classical N-body problem with point-mass sources is considered. This is a common and reasonable assumption: for a high number of commensurate interacting bodies, the use of shape-based mutual gravity would impact dramatically on the computational effort and would not provide relevant benefits to the accuracy of results, as discussed in section~\ref{sub:problemdef}. As detailed below, module~\ref{enum:mod-i} includes both direct N$^2$ integration and a parallel-GPU octree method. 
	
		\subsubsection*{The N-body problem}
			The equations of motion of the N-body problem are typically written by using Newton's law to compute the gravitational interactions between $N$ masses:
			\begin{equation}
				\label{eq:Newton}
				m_i \ddot{\textbf{R}}_i = G \sum\limits_{j=1, j \neq i}^N \frac{m_i m_j}{\lvert \lvert \textbf{R}_{ij} \rvert \rvert^3} \textbf{R}_{ij} \qquad \forall i = 1 :N
			\end{equation}
			where $\textbf{R}_{i}$ represents the position vector of the $i$-th body in an inertial frame, $m_i$ its mass, $G$ the universal gravitational constant and $\textbf{R}_{ij} = \textbf{R}_{j} - \textbf{R}_{i}$. A method implementing all N-to-N interactions is commonly referred as direct N$^2$ or particle-particle method~\citep{HE88, PH10}. Such method is very simple to implement and provides exact computation of the motion of the N-bodies without any physical approximation. The accuracy of the result depends solely on the numerical errors of the solver. The simple dynamics of direct N$^2$ method make it very easy to implement and parallelize. However, they are extremely inefficient from a computational point of view: $N(N-1)$ interactions must be computed for each body involved in the simulation and this leads to a time complexity of $O(N^2)$. A direct consequence is that such codes are limited in terms of maximum number of bodies to be handled. The bottlenecks are simulation time that becomes prohibitively long, and hardware capabilities with computer running out of memory, both effects occurring very rapidly as N increases. The direct N$^2$ method, even with advanced parallelism on processing units, is not suitable for efficient asteroid aggregation simulations that can involve more than few thousands of bodies.
	
		\subsubsection*{Hierarchical treecode algorithm}
			Algorithms based on tree data structures rely on more dynamic and adaptive computations that allow for a significant reduction of time complexity up to $O(N\ \log(N))$. The Barnes-Hut (also referenced as BH in the followings) algorithm~\citep{BH} groups particles using a hierarchy of cube structures. A node in the algorithm corresponds to a cube in physical space. Because of the use of octrees, each node has eight child nodes obtained by a simple homogeneous spatial subdivision performed along the three principal axis of the system. The tree is therefore built by recursive sub-division until each node of the tree contains zero or one particle. The structure is adaptive, implying that the size of the tree is not fixed but comes as a result of the repartition of the particles in the 3D space. The data structure can grow naturally to more levels in regions where the particle density is high. The octree is re-built at each time step~\footnote{See section~\ref{sub:problemdef} for more details on the choice of simulation time step(s)} of the dynamic simulation and consists of bodies stored at terminal nodes, called leaf nodes, and intermediary internal nodes that behave as clusters of particles. The algorithm uses the newly obtained adaptive octree data structure to compute the center of mass of the cubes involved in the simulation and the forces applied to each body: body-to-body and body-to-node interactions are considered.
			
			\begin{figure}
				\centering
				\includegraphics[width=0.5\columnwidth]{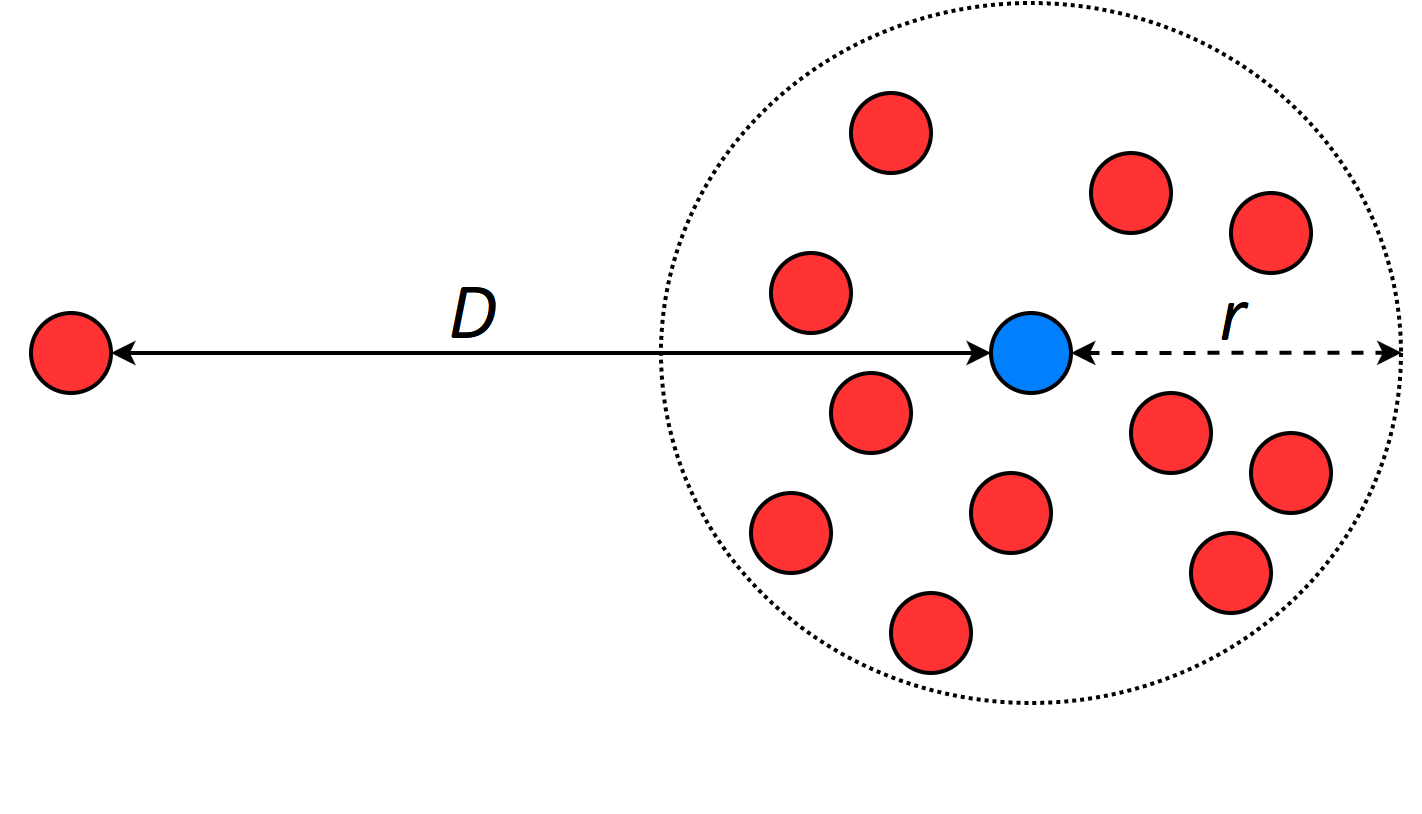}
				\caption{Schematics of Barnes-Hut clustering criteria}
				\label{fig:BH_criteria}
			\end{figure}
					
		\subsubsection*{Clustering}
			The algorithm relies on the idea that the force generated by a cluster of bodies can be approximated by treating the cluster as a single body. The accuracy of the approximations depends on the distance $D$ of the cluster from the body and the radius $r$ of the cluster of particles, as shown in Figure~\ref{fig:BH_criteria}. We can therefore define the accuracy through $\theta = r/D$, a user-tunable parameter used to set the error of the method. The position of center of mass and total mass of the node is computed for each leaf and internal node of the Barnes-Hut tree. For a given internal node $\mathcal{N}_{int}$:
			\begin{equation}
			\label{eq:COMinternal}
			\left\{
				\begin{array}{l}
			 		X_{\mathcal{N}_{int}} = \dfrac{1}{m_{\mathcal{N}_{int}}} \cdot \sum\limits_{i=1}^8  X_{child[i]} \cdot m _{child[i]} \\
		 			Y_{\mathcal{N}_{int}} = \dfrac{1}{m_{\mathcal{N}_{int}}} \cdot \sum\limits_{i=1}^8  Y_{child[i]} \cdot m _{child[i]} \\
		 			Z_{\mathcal{N}_{int}} = \dfrac{1}{m_{\mathcal{N}_{int}}} \cdot \sum\limits_{i=1}^8  Z_{child[i]} \cdot m _{child[i]} \\
		 			m_{\mathcal{N}_{int}} = \sum\limits_{i=1}^8 m _{child[i]}
				\end{array}
			\right.
			\end{equation}
			where ($X_{\mathcal{N}_{int}}, Y_{\mathcal{N}_{int}}, Z_{\mathcal{N}_{int}}$) are the coordinates of the position vector $\textbf{R}_{\mathcal{N}_{int}}$ along the three principal axes of the system, while $m _{child[i]}$ and ($X_{child[i]}$,  $Y_{child[i]}$, $Z_{child[i]}$) are respectively the mass and the coordinates of the position vector along the principal axes of the $i$-th child node of $\mathcal{N}_{int}$.
		
		\subsubsection*{Force computation}
			After the octree is built and all nodes contain the required information, parameter $\theta$ is computed for each body-node pair. ($\mathcal{B}$, $\mathcal{N}$):
			\begin{equation}
				\theta_{\mathcal{B},\mathcal{N}} = \dfrac{\text{Radius of} \ \mathcal{N} }{R_{\mathcal{N}}-R_{\mathcal{B}}}
				\label{eq:accuracy}
			\end{equation}
			This is compared with the value $\theta$ chosen by the user to set the accuracy and performance of the force estimation algorithm. A more detailed discussion on the effects of the user-tunable parameter $\theta$ on both computational time and accuracy of the algorithm is provided in section~\ref{sub:problemdef}. For each body $\mathcal{B}$, the tree is traversed from its root downwards along multiple branches and checks are performed at each encounter with a node $\mathcal{N}$, in order to compute force contributions:
			\begin{itemize}
				\item If $\mathcal{N}$ is a leaf node, the force contribution is evaluated as a classical body-to-body interaction:
				\begin{equation}
					F_{\mathcal{N} \rightarrow \mathcal{B}} = G \cdot m_{\mathcal{B}} \cdot m_{\mathcal{N}} \cdot(\dfrac{X_{\mathcal{N}}-X_{\mathcal{B}}}{r_{\mathcal{N}-\mathcal{B}}^3}, \dfrac{Y_{\mathcal{N}}-Y_{\mathcal{B}}}{r_{\mathcal{N}-\mathcal{B}}^3} , \dfrac{Z_{\mathcal{N}}-Z_{\mathcal{B}}}{r_{\mathcal{N}-\mathcal{B}}^3})
					\label{eq:Force_leaf}
				\end{equation}
				\item If $\mathcal{N}$ is an internal node and $\theta_{\mathcal{B},\mathcal{N}}<\theta$, the traversal is interrupted at this node and the force contribution is evaluated:
				\begin{equation}
					F_{\mathcal{N} \rightarrow \mathcal{B}} = G \cdot m_{\mathcal{B}} \cdot m_{\mathcal{N}} \cdot(\dfrac{X_{\mathcal{N}}-X_{\mathcal{B}}}{r^3}, \dfrac{Y_{\mathcal{N}}-Y_{\mathcal{B}}}{r^3} , \dfrac{Z_{\mathcal{N}}-Z_{\mathcal{B}}}{r^3})
					\label{eq:Force_thetaTrue}
				\end{equation}
				where r is the distance from the particle to the center of mass of the cube softened according to a parameter $\epsilon$:
				\begin{equation}
					r = \sqrt{(X_{\mathcal{N}}-X_{\mathcal{B}})^2 + (Y_{\mathcal{N}}-Y_{\mathcal{B}})^2 + (Z_{\mathcal{N}}-Z_{\mathcal{B}})^2 + {\epsilon}^2}
					\label{eq:Force_rComp}
				\end{equation}
				$\epsilon$ is called softening length and, as its name suggests, ensures the ``softening'' of gravitational forces when $\textbf{R}_{\mathcal{B}} \approx \textbf{R}_{\mathcal{N}}$, to avoid very high values of speed and acceleration that the integrator could not handle. Studies show that, without opting for a time-dependent adaptive softening parameter, $\epsilon$ must be a factor of at least twice the minimum distance between a body and the attractive cluster involved~\citep{Softening}. In the case of finite-size body interaction, with contact and collisions, the softening parameter is not needed, since no overlap between center of mass of a body and a node could occur.
				\item If $\mathcal{N}$ is an internal node and $\theta_{\mathcal{B},\mathcal{N}}>\theta$, the traversal continues along all eight child nodes of $\mathcal{N}$.
			\end{itemize}

		\subsubsection*{CUDA implementation of GPU-parallel Barnes-Hut}
			The implementation of the Barnes-Hut algorithm on a GPU using CUDA language is inspired by the work of Burtscher and Pingali~\citep{CUDA}. The physical domain is divided into sub-domains and the bodies are grouped following an octree structure. Similar to what done by~\citet{CUDA}, the numerical tasks to follow the Barnes-Hut algorithm have been divided among five kernels, which are written to minimize memory accesses. The tasks are executed sequentially on the GPU as follows:
			\begin{itemize}
				\item compute bounding box around all bodies (root of the octree)
				\item build the octree
				\item compute mass properties of each node
				\item sort bodies by distance (to speed up force evaluation)
				\item compute forces
			\end{itemize}
			Each body is assigned to a thread of the GPU, which traverses the octree to compute all force contributions acting on the body. Thanks to sorting, neighbor bodies require the traversal of a similar sub-part of the octree, whereas distant bodies may require completely different traversals.

	\subsection{Rigid-body dynamics and contact interactions}
	\label{sub:contact}
		Modules~\ref{enum:mod-iii} and~\ref{enum:mod-iv} deal with the creation of bodies and the definition of their inertial and surface properties. The latter depend on the contact method the user choose to use, among those available in~\ref{enum:mod-ii}. The following paragraphs recall briefly the main features of the code concerning body handling and contact interactions, focusing on the parameters relevant to each method. References are provided for a more complete discussion of each method.
		
		\subsubsection*{Creation of bodies and their properties}
			The shape of each body can be selected by the user among common geometries (sphere, box, cone,\dots) or directly input by the user as a triangulated mesh. The shape can also be generated as the convex envelope of a randomly created cloud of points. When dealing with a large number of bodies, methods exist to select their spatial distribution or pattern (regular grid, randomly uniform, Poisson-disk sampling,\dots) and bounding domain (common geometry or fitted in a triangulated mesh). Each body possesses 6 degrees of freedom, including both translational and rotational 3D motion. Random distribution routines can be used to set physical properties (size, density,\dots) and dynamical state (position, velocity, angular position, angular velocity) of bodies as well. Finally, the surface properties of each body can be defined. These are directly correlated to the contact method in use. Surface/contact parameters are discussed below both for smooth and for non-smooth methods. More details can be found in the documentation of~\cite{Chrono}.
		
		\subsubsection*{Collision detection}
			Collision detection is performed into two steps: a \textit{broad} and a \textit{narrow} phase. During the broad phase, pairs of bodies whose geometrical boundaries are close enough are identified. The bounding volume of a body is estimated based on its velocity and integration time step. If the bounding volume of two bodies overlap, the narrow phase is performed and contact points are precisely found using a GJK algorithm~\citep{Tasora2010,Fabio}.
			
		\subsubsection*{Contact method: smooth contacts}
			The Smooth Contact Method (SMC) implemented in C::E~\citep{Fleischmann2015} is a force-based soft-body DEM method. The equations of motion are formulated as a system of Differential Algebraic Equations (DAE): namely a system of Ordinary Differential Equations (ODE) to reproduce the dynamics and Algebraic Equations (AE) for the kinematic constraints.
			Forces are exchanged between bodies at contact using a two-way normal-tangent spring-dashpot system. The constitutive model of the spring-dashpot system can be selected between the simple Hooke and the more realistic Hertzian model. Relevant parameters of the SMC are coefficients of friction (static, dynamic, spinning), cohesion (as an attractive force at contact points), stiffness and damping. As any soft-body DEM, it is suitable for smooth problems with no discontinuities but requires very small time steps for cases with high stiffness to avoid numerical instability.
			
		\subsubsection*{Contact method: non-smooth contacts}
			The Non-Smooth Contact (NSC) method implemented in C::E~\citep{Tasora2010,AnitescuTasora2010,TasoraAnitescu2011} is an impulse-based hard-body method. Unlike the SMC, the equations of motion are formulated as Differential Variational Inequalities (DVI) and require the solution of a Cone Complementarity Problem (CCP) at each time step. The NSC shares with the SMC parameters defining friction (static, dynamic, spinning) and adhesion. The handling of contact interaction is however much simpler and solely relies on the restitution coefficient, defined as the ratio between velocity after and before the collision. Due to its impulse-momentum formulation, the NSC is best suited for problems with discontinuities or with nearly rigid contacts (high stiffness).
						
		\subsubsection*{Contact method: non-smooth contacts with compliance}
			In the context of non-smooth dynamics, a method is available to simulate contacts with compliance and damping~\citep{Tasora2013}. As for the NSC method, the equations of motion are formulated as DVI and are based on a impulse-momentum formulation. However, compliance and damping are enforced at the constraint level and the method is suitable not only for non-smooth problems but also to simulate the elastic and smooth behavior typical of soft-body DEM models. Also, since it does not rely on the solution of a DAE, but its solution is found after solving a CCP-based DVI, this method does not require the very small time step required by soft-body DEM. Its parameters are analogous to that of SMC: friction (static, dynamic, spinning), cohesion, stiffness and damping.

	\subsection{Numerical solvers}
	\label{sub:numerical_solvers}
		As discussed in the previous section, when using the SMC approach the dynamics are written as a system of DAE, and when using the NSC approach the dynamics are written as a DVI problem. From the numerical point of view, DAE require to numerically integrate ODEs, while DVI require the solution of an optimization problem at each time step. Despite the radical difference in approaching the solution, both approaches require a time stepper and a nonlinear solver. A large variety of time steppers and solvers is available in C::E~\citep{TasoraIntegrationCE}: the most interesting for gravitational-granular dynamics applications include symplectic methods (semi-implicit Euler, leapfrog) and Runge-Kutta methods (RK45, explicit Euler, implicit Euler, trapezoidal, Heun), to be complemented with either iterative or direct solvers. In particular, symplectic methods are very suitable for gravitational dynamics problem and semi-implicit Euler method provide the best performance among them, when paired with an iterative solver~\citep{AnitescuTasora2010,Mangoni2018}. This is suitable and effective to solve both DAE and DVI problems.
		
	\subsection{Post-processing: aggregate identification}
	\label{sub:shape}
		Module~\ref{enum:mod-viii} contains routines for the post-simulation analysis of results. We provide here one example that applies to the case of rubble-pile aggregation simulations. One of the tasks to be performed after such simulations is to identify the aggregate(s) of bodies and determine their properties such as shape, mass, inertia and void fraction (porosity).
	
		\subsubsection*{Aggregate identification}
\begin{figure}
				\begin{center}
					\includegraphics[width=0.95\columnwidth]{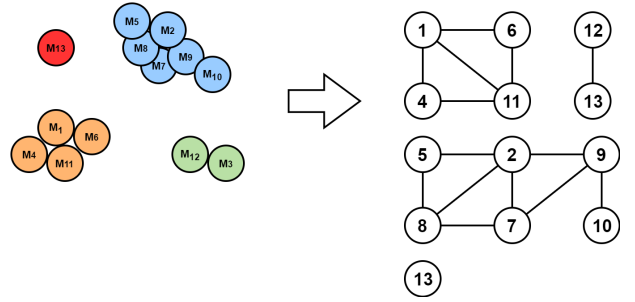}
					\caption{Example of global contact graph construction}
					\label{fig:Ex-GlobalGraph}
				\end{center}
			\end{figure}

			The single or multiple aggregates obtained are identified through a method based on graph theory~\citep{Graph}. First, the graph $\mathscr{G}$ is initialized to represent the whole system. Bodies are vertices of $\mathscr{G}$ and the size of the vertex set $E(\mathscr{G})$ is $N$ (number of bodies). Edges of $\mathscr{G}$ represents physical contacts between objects: two bodies $M_i$ and $M_j$, represented by vertices $P_i$ and $P_j$ of $\mathscr{G}$, are in contact if and only if there is an edge $\{P_i, P_j\}$ linking $P_i$ to $P_j$. At the end of the simulation, a contact container is filled with pairs $\text{ID}_i, \text{ID}_j$ corresponding to pairs of bodies in contact with each other. The graph $\mathscr{G}$ is filled with all existing contacts, drawn as edges in the graph. A schematic example is shown in Figure~\ref{fig:Ex-GlobalGraph}.
	
			Aggregates obtained correspond to the connected subgraphs of $\mathscr{G}$. Single bodies, that do not belong to any aggregate (e.g because they have escaped from the re-aggregation processes) are represented by isolated vertices. The number of aggregates is defined as the number of connected components of $\mathscr{G}$ with a size greater than one, that is the number of the subset made of the connected subgraphs of $\mathscr{G}$ with more than one vertex. A Depth First Search (DFS) algorithm~\citep{Graph} is used to provide an effective way of traversing simple graphs. The output is a classification of the edges and a spanning tree that can be used for finding the connected components of a graph. Using such algorithm results in a complexity of  $O(N + E)$ (where $N$ and $E$ are respectively the number of vertices and edges of $\mathscr{G}$) for the differentiation process.
	
		\subsubsection*{Shape of the aggregate}
			The problem of defining the shape of a body made up of several discrete elements is not trivial and has a non-unique solution: infinite surfaces exist to envelope a given set of points. After identifying the aggregate and its children bodies, we address the problem of defining its overall shape by means of an alpha-shape algorithm~\citep{alphaShapes}. This computes the surface by enveloping a given set of points: in our case all vertices of children bodies. Unlike the simpler convex hull envelope, the alpha-shape surface is, in general, non-convex. The working principle of the alpha-shape algorithm can be visually represented as follows. We start from a cloud of points (vertices of children bodies). Let us imagine to roll a sphere of radius $R_\alpha$ over the cloud of points. The alpha-shape envelope is the surface created by the sphere as it rolls over the cloud of points. In the limiting case where $R_\alpha=\infty$, the surface enveloping the cloud would be a convex hull. For a finite value of $R_\alpha$, non-convex envelopes can be obtained. Typically, $R_\alpha$ should be on the same order of magnitude of the mean distance between vertices. As mentioned, the problem has a non-unique solution and the choice of $R_\alpha$ makes the process arbitrary. This is affecting in a relevant manner the computation of global properties of the aggregate such as volume, void fraction (porosity) and shape-related quantities (elongation, axis ratios, \dots). To solve for this arbitrariness, we define an admissible range of enveloping surfaces, within a range of limiting cases. In general, two limiting surfaces can be defined: (\textit{i}) the minimum volume surface, which is found using the minimum value of $R_\alpha$ such to have no surfaces delimiting cavities inside the aggregate, and (\textit{ii}) the maximum volume surface, which is the convex hull of the aggregate, obtained with $R_\alpha=\infty$. Global properties are then computed considering the uncertainty on their values due to the arbitrary definition of the surface, and range between values computed using minimum and maximum volume surfaces.

\section{Validation scenarios}	
\label{sec:validation}
	Original modules of C::E dealing with the modeling of granular dynamics have been widely tested and validated against benchmark and experimental tests~(see ``Validation Studies section in~\cite{Chrono},~\cite{HeynPhD} and others). The results of further test simulations are shown here to validate the gravity modules and their integration into the overall implementation. In particular, the accuracy of direct N$^2$ and octree gravity models are assessed and their performance is compared in terms of computational time. Conservation of energy, linear and angular momentum are checked for the purely gravitational case and for the coupled gravity-collision problem.
	
	\subsection{Direct N-body integration}
\begin{figure}
			\begin{center}
				\includegraphics[width=0.99\columnwidth]{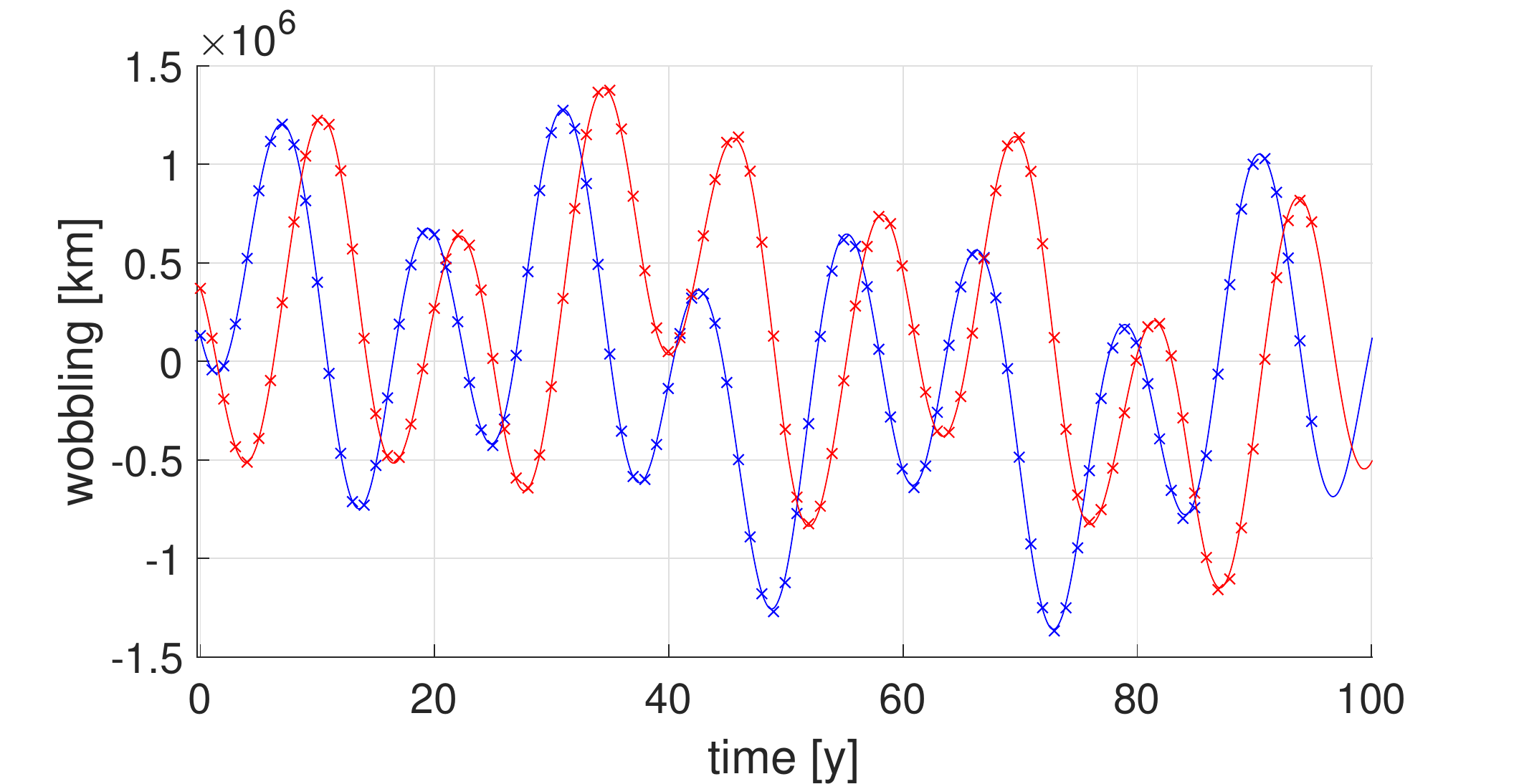}
				\caption[]{Validation scenario: in-plane (x and y components in the ecliptic J2000 plane) wobbling of the Sun around the barycenter of the Solar System during a 100-year timespan. Comparison between JPL's Horizons ephemerides\footnotemark (solid line) and numerical integration (cross markers)}
				\label{fig:SunWobb}
			\end{center}
		\end{figure}

		The first set of tests concerns the accuracy of force computation for the direct N$^2$ method. Unlike the octree method, direct integration considers all gravitational interactions with no physical approximation. We performed several simulations of direct integration between mutually interacting bodies and in particular:
		\begin{itemize}
			\item conic solutions of the two-body problem are reproduced as function of orbital energy;
			\item orbits of planets in the Solar System are propagated for two-thousand years;
			\item Halo and Lyapunov orbits around collinear libration points of the Sun-Earth and Earth-Moon circular restricted three-body problems (CR3BP) are reproduced.
		\end{itemize}
		When compared to the expected analytical (two-body problem) or numerical solution (N-body problem and CR3BP), the code is able to reproduce \textit{exactly} the result, where the term \textit{exactly} (here and from now on) implies to within round-off accuracy or numerical integration truncation errors. As mentioned in section~\ref{sub:numerical_solvers}, the semi-implicit Euler method is a very effective choice to solve gravitational dynamics, thanks to its symplectic nature that guarantees conservation of energy within the system. Energy, linear and angular momentum are conserved \textit{exactly} for all cases, consistently with the nature of the direct N$^2$ formulation that introduces no approximation errors in the  model. The successful outcome of above tests guarantee that gravitational forces are computed correctly for each body. Figure~\ref{fig:SunWobb} shows an example related to one of the aforementioned scenarios. The expected in-plane wobbling of the Sun around the barycenter of the Solar System (data from JPL's Horizons ephemerides, solid line) is compared and matches well with the outcome of a direct N$^2$ integration of the motion of Sun and planets during a 100-year timespan (cross markers).
		
		\footnotetext{\url{https://ssd.jpl.nasa.gov/}}

	\subsection{GPU-parallel octree}
		We evaluate here the performance of the GPU-parallel implementation and compare it to direct N$^2$ integration in terms of computational time and accuracy. Results presented here were obtained using a medium-range laptop with
		\begin{itemize}
			\item Intel(R) Core(TM) i7-6500U CPU @3.1Ghz
			\item Nvidia Geforce 940M (384 CUDA cores).
		\end{itemize}
		Although certainly dependent on hardware architecture, the results here shown identify trends and behaviors that are not dependent on hardware, but only on the algorithms used. To assess the performance of algorithms, two different sets of simulations are performed. For each simulation, bodies are created at time zero in random positions within a cubic domain and with a uniform distribution. Collision detection and interaction is disabled during the entire duration of the simulation, such that bodies only experience mutual gravity. The results of validation scenarios are in agreement with previous assessments concerning tree codes~\citep{BH,Hernquist1987}.
	
		\subsubsection*{Computational time}
\begin{figure}
				\begin{center}
					\includegraphics[width=0.9\columnwidth]{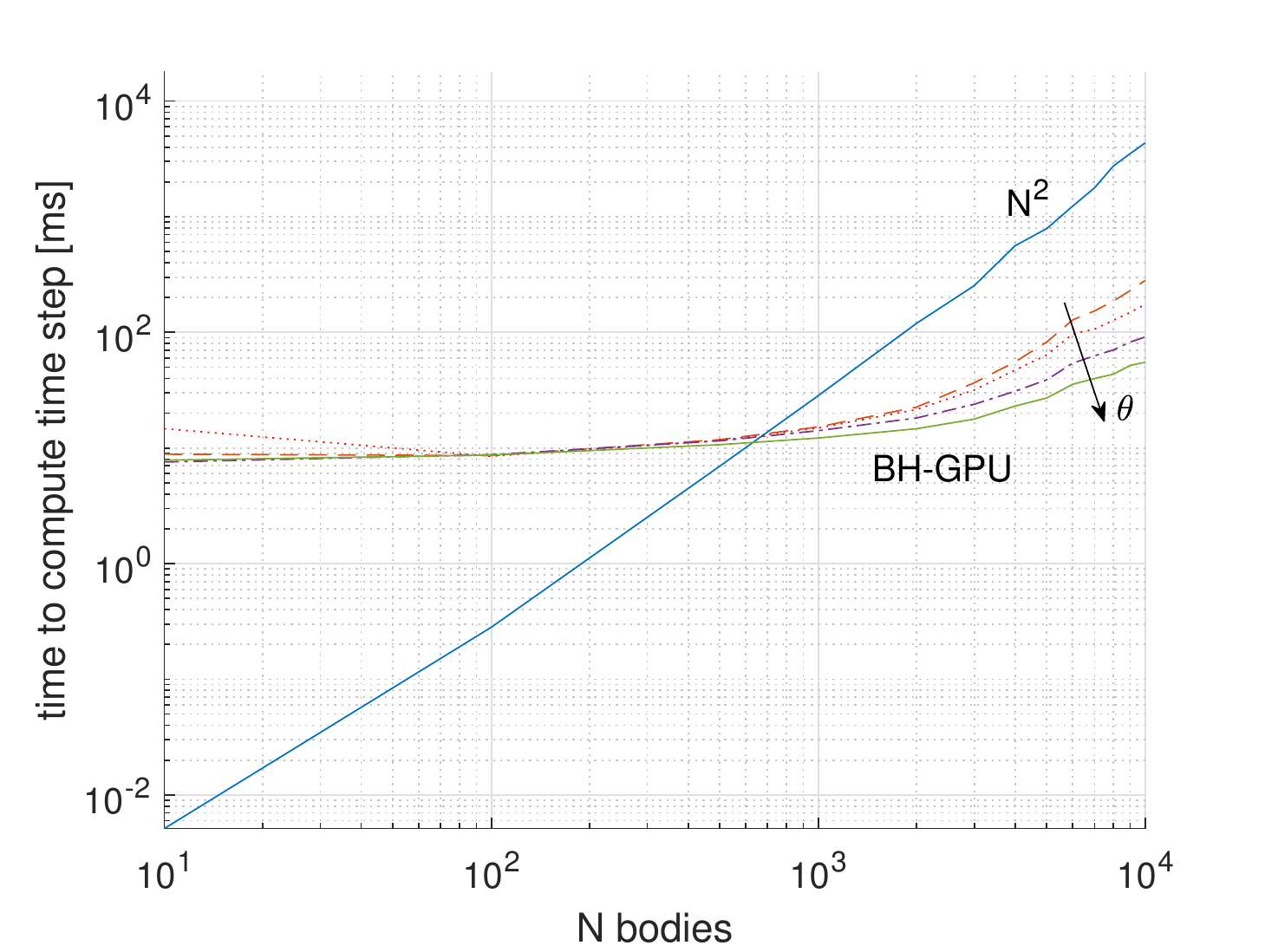}
					\caption{Time to compute one integration time step as function of number of bodies in the simulation. Results are shown for direct N$^2$ integration vs Barnes-Hut GPU-parallel algorithm ($\theta=0.1,0.25,0.5,0.75$)}
					\label{fig:tdt}
				\end{center}
			\end{figure}

			A first set of simulations is arranged to compare the computational time required by each algorithm. Simulations are performed for several scenarios, including $N$ ranging from 10 up to 10$^4$, and using both direct N$^2$ integration and Barnes-Hut GPU algorithm with different level of accuracy ($\theta=0.1,0.25,0.5,0.75$). We compare the time required by the code to compute one integration time step. In principle, results can be compared after computing only one time step. However, to avoid any bias and for the sake of a more accurate profiling, we integrate the dynamics forward for several time steps and take the average time required to compute one of them. Also, we want the dynamics to be steady such to not interfere with the evaluation of performance. To this goal we use a very short time step (10 s) and a total duration of the simulation of 10$^4$ s. Results are shown in Figure~\ref{fig:tdt} using a logarithmic scale. As expected, the cost of direct integration increases with N$^2$. As discussed in section~\ref{sub:gravity}, the BH-GPU algorithm executes a sequence of kernels and the largest amount of computational time is shared between the creation of the octree and the evaluation of the gravitational body-body or body-cluster interactions. Both octree and gravity evaluation depend on the number of bodies in the system, but to very different extents. Figure~\ref{fig:tdt} shows clearly that for a low number of bodies, the computational time required is dominated by the octree creation phase, which is independent from the choice of the accuracy parameter $\theta$ and increases very slowly with N. Conversely, for high number of bodies the time expense is dominated by the evaluation of the gravitational interactions, which instead depends on $\theta$. When comparing direct N$^2$ with BH-GPU, it clearly appears that for a low number of bodies (approximately N<700, the same for which $\theta$ is not affecting the time expense of BH-GPU), the process of building the octree makes the BH-GPU algorithm much more expensive than direct N$^2$ integration. Conversely, for a high number of bodies, the BH-GPU is much faster than direct N$^2$. For N=1,000 the BH-GPU turns to be nearly two times faster and for N=10$^4$ it is faster of more than one order of magnitude. For a further increased number of bodies, direct N$^2$ simulations become practically unfeasible and BH-GPU becomes the only feasible option. As discussed, BH-GPU algorithm can be tuned in accuracy: this affects the computational time as shown for different $\theta$ values used in Figure~\ref{fig:tdt}. In this high-N regime, the computational time of BH-GPU ranges from $O(N^2)$ (with $\theta=0$, i.e. no clustering of bodies, equivalent to direct integration) up to $O(N\ \log(N))$ as $\theta$ increases.
	
			\begin{figure}
				\begin{center}
					\includegraphics[width=0.9\columnwidth]{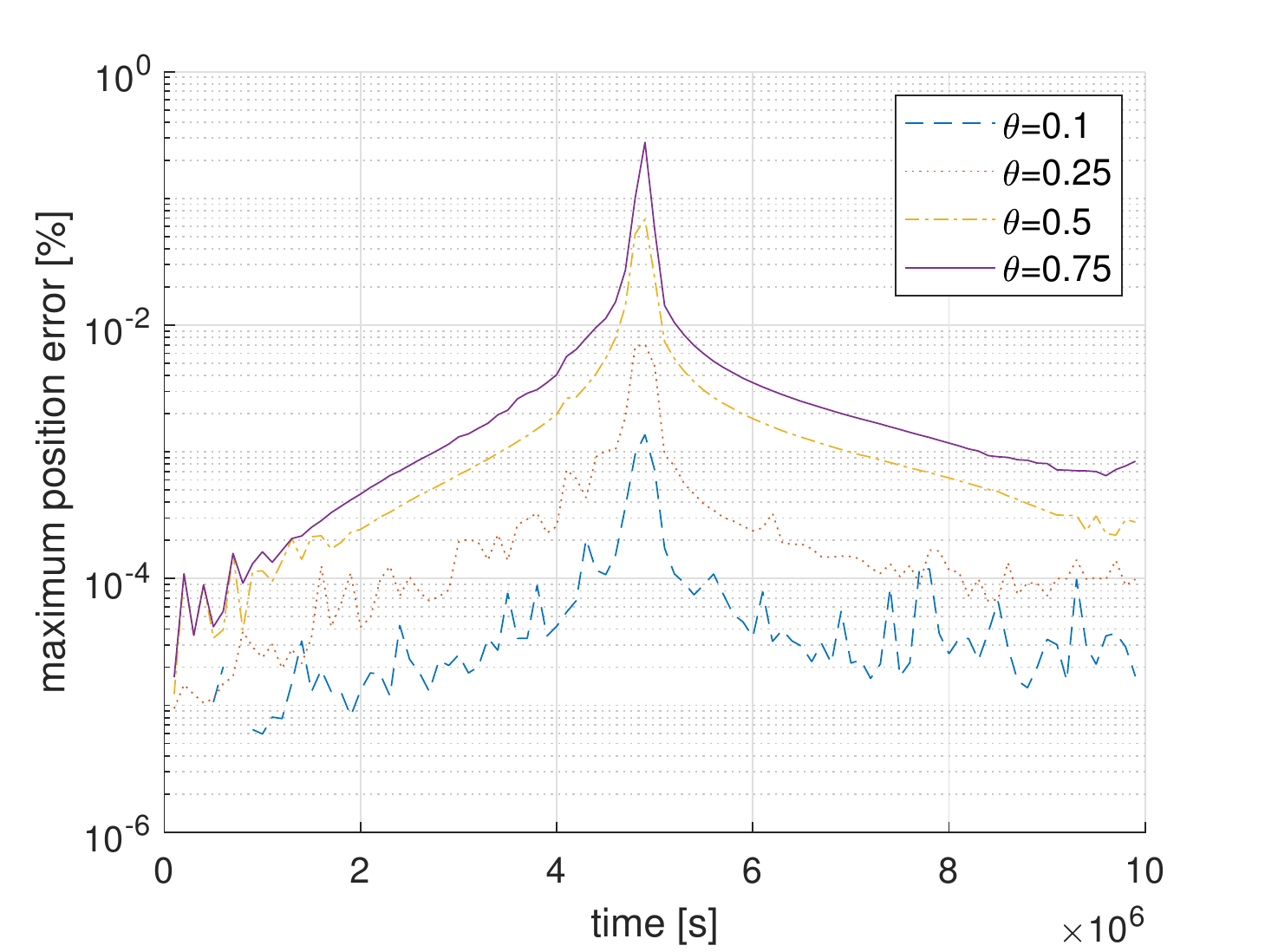}
					\caption{Maximum relative position error in time. The error is computed for simulations of 1,000 bodies with Barnes-Hut GPU-parallel algorithm ($\theta=0.1,0.25,0.5,0.75$), and it is relative to a simulation with direct N$^2$ integration}
					\label{fig:error}
				\end{center}
			\end{figure}
	
		\subsubsection*{Accuracy}
			A second set of simulations is used to assess the accuracy of the BH-GPU algorithm. In this case, we wait for the bodies to complete a full orbit around the barycenter of the system and come back to their initial state at rest. Since collisions are disabled, all bodies are undisturbed in their gravity-driven path and the system exhibits a nearly perfect energy-conserved pulsating contracting-dilating behavior. This simulation time frame gives us a commensurate estimate of typical time duration of aggregation phenomena and provides us with every possible gravity-related condition, i.e. both, when bodies are far away and when they are close to each other. Compared to the first set of simulations, we use a larger time step and we integrate forward the dynamics for a much longer time. Results are here shown for the case of N=1,000 and for $\theta=0.1,0.25,0.5,0.75$. A simulation using direct N$^2$ integration is also performed and used as a real-world result to assess the accuracy of the BH-GPU algorithm. At each time step, the position of each body in the system is compared to its corresponding real-world one and the relative error is evaluated as percentage of the latter
			\begin{equation}
				e_i(t)=\frac{\|\mathbf{r}_{i_\text{BH}}(t)-\mathbf{r}_{i_{N^2}}(t)\|}{\|\mathbf{r}_{i_{N^2}}(t)\|} \cdot 100
			\end{equation}
			where $\mathbf{r}_{i_\text{BH}}(t)$ and $\mathbf{r}_{i_{N^2}}(t)$ are the positions of body $i$ at time $t$ when using, respectively, BH-GPU and N$^2$ algorithms. Figure~\ref{fig:error} shows the maximum position error among bodies in the system at each time step of the simulation. As expected, the error increases monotonically with $\theta$, which is a direct mean and user-tunable parameter to select the desired level of accuracy. Few more features are worth of discussion. The error is shown to be time-dependent and exhibit a peak at half simulation time, i.e. at the end of contraction phase. This is expected, as the relative position error depends on the magnitude of accelerations acting on the bodies. The higher the acceleration, the higher the relative error when evaluating body-body vs body-cluster interactions. Accordingly, the error is shown to be higher in the middle part of the simulation, where bodies are closer to each other and it increases during contraction and decreases as they get further to each other. It is interesting to look at the worst case depicted in Figure~\ref{fig:error}. At the end of contraction phase, the error between the two extremes $\theta=0.1$ and $\theta=0.75$ is of about two orders of magnitude. However, the maximum relative error obtained using $\theta=0.75$ is still quite small, in the order of 0.3\%. Also, it is worth noting that this simulation provides us with an extremely conservative estimate of the accuracy. This is because real-world scenarios include the dissipative effects introduced by collisions and often entail only a part of the contraction phase simulated here. For example, if collisions were enabled in the scenario considered here, they would become very relevant starting at approximately 2.5$\cdot$10$^6$~s (\textit{transient phase}) and the bodies would never get as close to each other as they are in the [3.5, 6.5]$\cdot$10$^6$~s range due to their physical extent (\textit{aggregate phase}). They will therefore experience lower gravitational acceleration actions and position errors due to BH-GPU clustering: surely lower than the worst-case 0.3\% value found here.

	\subsection{Collisions}
\begin{figure}
			\begin{center}
				\includegraphics[width=0.9\columnwidth]{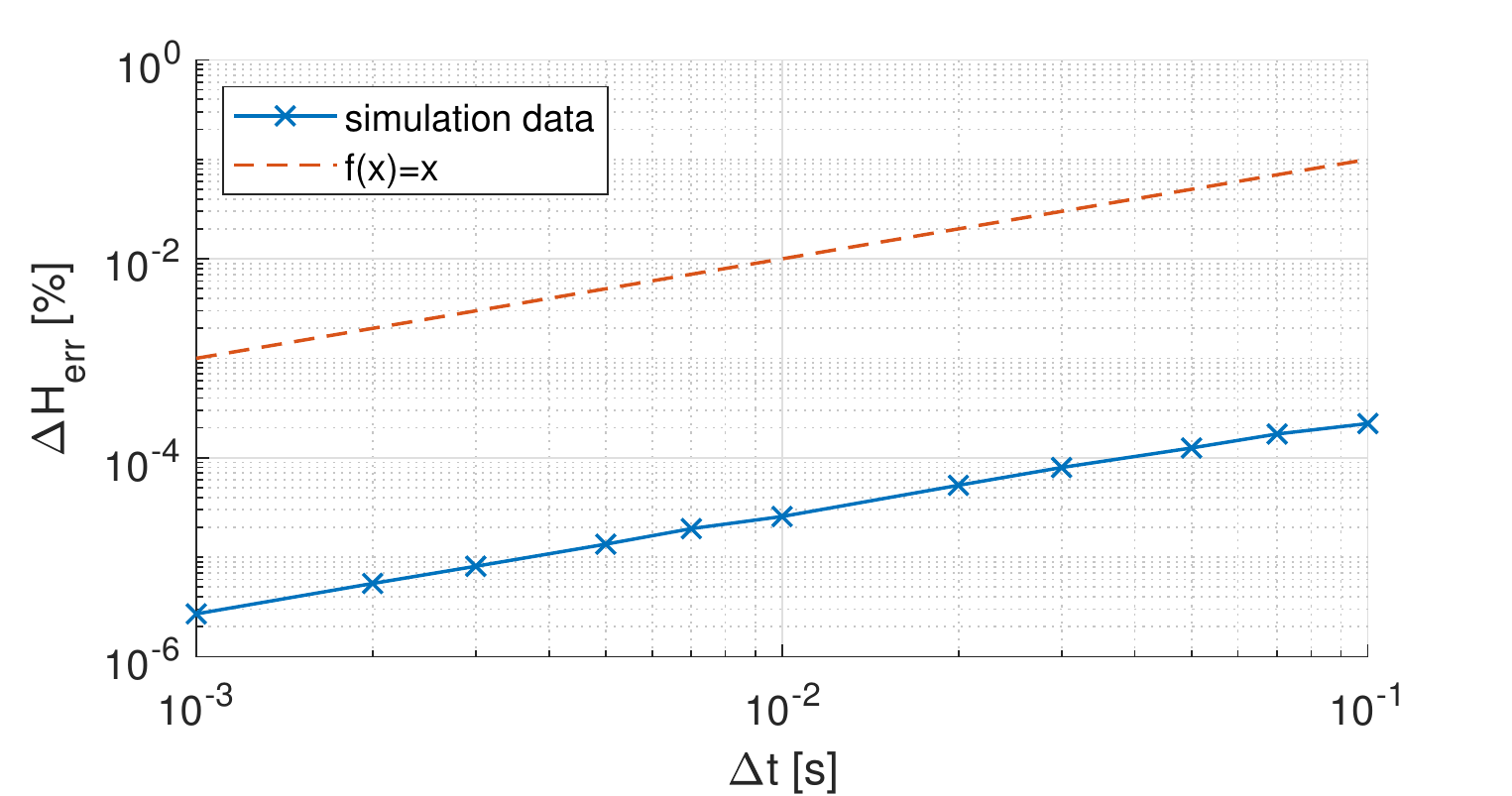}
				\caption{Error on total angular momentum of the system after one simulation time step, with both gravity and contact interactions. Error is shown in percentage, normalized to the initial value of angular momentum. The linear trend of the error as function of time step duration is highlighted.}
				\label{fig:H_dt}
			\end{center}
		\end{figure}

		A set of simulations is performed to test conservation laws during collisions and contact interactions between bodies. In this case, we simulated the settling dynamics of an already formed self-gravitating aggregate with $\sim$1,000 convex-hull bodies. The dynamics are propagated forward for one time step and contacts are solved using both smooth and non-smooth methods. The difference of physical and dynamical quantities are monitored. Several simulations are performed using the SMC method and different time steps, ranging between 10$^{-3}$ and 10$^{-1}$ seconds. The same is done using the NSC method which is able to handle higher time steps, between 10$^{-1}$ to 10 seconds. A more detailed discussion on time step selection criteria is reported in Section~\ref{sub:problemdef}. Unlike the purely gravitational case, energy, angular and linear momentum are not conserved \textit{exactly} and collisions introduce approximation errors in the models. Since the semi-implicit Euler method is a first order integrator, the error is expected to depend linearly on the time step. This is confirmed by our simulations, as shown in Figure~\ref{fig:H_dt}, which refer to the SMC case. It shows the angular momentum error in percentage, normalized to its initial value, for different time steps. As expected, a clear linear trend is identified for the error, as it scales with the time step.
					
\section{The asteroid aggregation problem}
\label{sec:asteroid}
	This section discusses how rubble-pile scenarios are simulated using the code described above (section~\ref{sec:code}). We discuss the appropriate choice of methods to use, how to tune their parameters and how to set up the numerical simulation.
	
	\subsection{Addressing the problem}
	\label{sub:problemdef}
		To properly address the problem of simulating rubble-pile dynamics, we study the fundamental interactions that occur between bodies during gravitational aggregation phenomena. These can be associated to phases of the gravitational aggregation process:
		\begin{itemize}
			\item \textit{gravitational phase}: the dynamics of the bodies is only driven by mutual gravity and they do not experience contact interactions.
			\item \textit{transient phase}: this phase begins as bodies start to collide with each other and form small aggregates of few bodies each. This phase is characterized by an extremely chaotic motion of the bodies that are simultaneously subjected to both gravity and random collisions. From a macroscopic point of view, the overall dynamics of the system is dominated by mutual gravity between bodies and newly forming aggregates. However, at particle level, the dynamics is dominated by collisions, which creates impulsive (non-smooth contact) or quasi-impulsive (smooth) forces on a much faster scale than gravitational one. The combined effects at system and body level, and the dissipative nature of collisions leads eventually the system to a steady state condition.
			\item \textit{aggregate phase}: after transient phase, the system reaches a steady state equilibrium where bodies are either clustered in one or more aggregates, or dispersed.
		\end{itemize}
		The most challenging phase to be simulated is certainly the transient phase, where both gravitational and collision dynamics play a major role. Since gravity and collisions act at different scales, we explore the capability and suitability of the code by studying such dynamics separately. This leads to meaningful and general conclusions on how to numerically simulate such interactions, also applicable to the fully coupled problem.
		
		\subsubsection*{Gravitational dynamics}
			In this early stage of the aggregation process, bodies are not in contact with each other and only interact gravitationally. As discussed in Section~\ref{sec:validation}, the choice of the most suitable method depends on the number of bodies involved: direct N-body for few bodies, octree for a large number of bodies with a hardware-dependent threshold typically between 1,000 and 10,000 bodies. In order to correctly simulate the N-body dynamics, the integration time step $\Delta t_g$ must be consistent to the characteristic time $T_g$ of the problem and must sample the dynamics at least twice per $T_g$. In particular, the following must be satisfied~\citep{Fabio}:
			\begin{equation}
			\label{eq:dt_g}
				\Delta t_g<\frac{T_g}{2}=\frac{1}{2\sqrt{G\rho}}
			\end{equation}
			where $G$ is the universal gravitational constant and $\rho$ is the material density of the bodies. For typical values of asteroid material density ranging between 1 and 4 g/cm$^3$~\citep{Richardson}, maximum time steps are in the order of thousands of seconds.

		\subsubsection*{Collisional dynamics}
			As done for gravitational dynamics, we study here the fundamental features related to collisional interactions between bodies. We analyze the different phases and numerical tasks involved when solving for collisions and derive numerical constraints. From the numerical and algorithmic point of view, collisional dynamics requires the solution of two different problems: collision detection and collision output determination.
			
			The collision detection algorithm operates at each time step to identify contacts between body pairs. To ensure proper collision detection, the time step must be adequate: if the time step is too high, a collision could be missed. To quantify the dynamical constraint on the time step, we consider here the case of a collision between two spheres of radius $R$. In particular, we consider the limiting case when a collision is not detected. It is easy to assess that, for the case of a direct collision occurring along the line connecting the centers of the spheres, the distance traveled during the time step $\Delta t_d$ must be lower than 4$R$, and then
			\begin{equation}
			\label{eq:direct_coll}
				\Delta t_d<\frac{4R}{v}
			\end{equation}
			where $v$ is the relative velocity between bodies. Eq.~\eqref{eq:direct_coll} can be generalized for the case of grazing collisions~\citep{Sanchez2011}:
			\begin{equation}
			\label{eq:grazing_coll}
				\Delta t_d<\frac{2\sqrt{4R\delta-\delta^2}}{v}
			\end{equation}
			where $\delta$ is the maximum overlap allowed between bodies: the smaller the overlap allowed, the smaller the time step required and the more precise the detection algorithm. Note that the case of direct collision in \eqref{eq:direct_coll} can be retrieved from \eqref{eq:grazing_coll} when $\delta=2R$ (complete overlapping). When considering a system of $N$ bodies, $v$ is the maximum relative velocity in the system and $2R$ is the sum of the radii of the two smallest particles. Also, although \eqref{eq:grazing_coll} refers to a simplified case with spheres, the same relation can be conservatively used with convex hulls of any shape, where $R$ represents their minimum characteristic size.
			
			The second task is to solve for the actual collisional dynamics when bodies are in contact, in order to compute the collision output. This task is performed by the contact method selected, as discussed in section~\ref{sub:contact}. Hard- and soft-body methods deal with a very different modeling of the physics at contact. Hard-body methods model instantaneous collisions and deal with non-smooth dynamics based on impulse-momentum formulation. Since the collision is instantaneous, it does not make sense to speak of characteristic time involved in the dynamical process and then to derive a requirement on the time step from that. However, the time step still plays a major role and has to be selected properly to ensure the stability of the numerical solver. Hence, unlike gravity, collision detection and, as discussed below, soft-body methods, the choice of the time steps for hard-body methods is not driven by dynamical constraints, but by the properties of the numerical solver. We consider here the case of the semi-implicit Euler method. Looking at its stability region~\citep{Niiranen1999}, we can derive a constraint on the time step to be used, as function of the roots of the characteristic equation. However, our problem involves many bodies and requires the solution of a CCP at each time step: the roots of the characteristic equation of such problem are not of easy determination. Alternatively, an empirical but very effective method is to determine a proper time step by comparison against known benchmark problems or known behavior of the system. For example, when benchmark problems are not available, a very convenient way is to tune the time step according to a desired level of accuracy in terms of conservation of total angular momentum of the system $H_{\text{tot}}$. As discussed in Section~\ref{sec:validation}, the error depends linearly on the time step. This linear dependency provides us with a very effective means to estimate and select properly the time step. Further examples on how the time step can be set are provided in C::E technical documentation and validation studies~\citep{Chrono}, with reference to non-smooth (NSC) and non-smooth visco-elastic methods.
			
			On the other hand, unlike hard-body ones, soft-body methods directly rely on the physical modeling of collisional dynamics. The collision occurs in a finite time and foresee the visco-elastic behavior of the bodies at contact. Collisions are modeled by means of a spring-dashpot system at the contact point. We can therefore identify a characteristic time related to the fundamental frequency of the spring-dashpot system $\omega_k$~\citep{Tsuji1993}:
			\begin{equation}
				T_k=\frac{2\pi}{\omega_k}=2\pi \sqrt{\frac{m_r}{k}}
			\end{equation}
			When considering a system of $N$ bodies, $\omega_k$ is the highest frequency in the system, i.e. made of the combination of highest stiffness $k$ and lowest reduced mass $m_r$\footnote{$m_r$ is the reduced mass of the two least massive bodies in the system}. Accordingly, the time step must be at least two times smaller than the characteristic time
			\begin{equation}
			\label{eq:dt_k}
				\Delta t_k<\frac{T_k}{2}=\frac{\pi}{\omega_k}
			\end{equation}
			Practical applications typically make use of smaller time steps, ranging from $0.1 T_k$ to $T_k/\pi$~\citep{Sanchez2011}, or even smaller than $0.01 T_k$~\citep{Herrmann1998}.
			
			To summarize, two different constraints on the simulation time step can be derived after the analysis of the collisional process. The first one is expressed by Eq.~\eqref{eq:grazing_coll}. It concerns the capability of the code to properly detect collisions and does not depend on the contact method in use. The second one depends on the contact method in use. In case of impulsive contacts, it depends on the stability of the numerical solver and can be determined empirically. In case of smooth dynamics, it depends on the dynamics of contact as expressed in Eq.~\eqref{eq:dt_k}. Both requirements shall be satisfied to properly simulate collisional processes. This duality is easily resolved by enforcing the most stringent requirement, i.e. by choosing the lowest time step required. As a general rule, collision detection will be the bottleneck for high velocity impacts and vice versa, an accurate collision output will be the driving criteria in case of low velocity collisions. In the attempt to quantify in a more rigorous way this general criterion, we investigate possible links between Eq.~\eqref{eq:grazing_coll} and~\eqref{eq:dt_k}. Indeed, while the two effects appear not to be directly coupled for hard-body methods, some considerations can be made for the case of soft-body methods, where both~\eqref{eq:grazing_coll} and~\eqref{eq:dt_k} are built upon the dynamics of the problem. In particular, it is reasonable to assume that the maximum overlap allowed during collision detection $\delta$ must be consistently smaller than the maximum $\Delta x$ contraction of the spring-dashpot system. If this is not satisfied, the visco-elastic behavior at contact will not be accurately resolved. The maximum contraction $\Delta x$ of a spring-mass system can be computed based on energy conservation consideration, by equating the total energy right before contraction (only kinetic energy, no contraction) to the total energy at maximum contraction (only potential energy, zero velocity):
			\begin{equation}
				\frac{1}{2}m_r v^2=\frac{1}{2}k\Delta x_s^2
			\end{equation}
			from which is easy to find that
			\begin{equation}
				\Delta x_s=\frac{v}{\omega_k}
			\end{equation}
			In the case of a spring-dashpot system, the total energy is no longer conserved and a fraction of the initial kinetic energy is dissipated. The maximum contraction will be lower in this case, reduced by a factor $\xi$<1
			\begin{equation}
				\Delta x_{sd}=\xi\frac{v}{\omega_k}
			\end{equation}
			Now we define the maximum overlap allowed during collision detection $\delta$ as a small fraction of $\Delta x_{sd}$:
			\begin{equation}
				\delta=\eta\xi\frac{v}{\omega_k}
			\end{equation}
			with $\eta<<1$, or equivalently
			\begin{equation}
			\label{eq:max_delta}
				\delta=\varepsilon\frac{v}{\omega_k}
			\end{equation}
			with $\varepsilon=\eta\xi<<1$. We can now substitute~\eqref{eq:max_delta} into~\eqref{eq:grazing_coll} and get
			\begin{equation}
			\label{eq:dt_c_k}
				\Delta t_d < \frac{2\sqrt{4R\varepsilon\frac{v}{\omega_k}-\varepsilon^2\frac{v^2}{\omega_k^2}}}{v}
			\end{equation}
			This relation correlates the two time steps involved in the numerical simulation of collision processes. In particular, we compare~\eqref{eq:dt_c_k} with~\eqref{eq:dt_k}. The collision detection time step $\Delta t_d$ and the stiffness time step $\Delta t_k$ are equal if
			\begin{equation}
			\frac{2\sqrt{4R\varepsilon\frac{v}{\omega_k}-\varepsilon^2\frac{v^2}{\omega_k^2}}}{v} = \frac{\pi}{\omega_k}
			\end{equation}
			which can be rewritten as
			\begin{equation}
				4R\varepsilon\frac{v}{\omega_k}-\varepsilon^2\frac{v^2}{\omega_k^2}=\frac{\pi^2}{4}\frac{v^2}{\omega_k^2}
			\end{equation}
			and finally, after trivial algebraic manipulation, as
			\begin{equation}
				v=\frac{4R\varepsilon}{\frac{\pi^2}{4}+\varepsilon^2}\omega_k
			\end{equation}
			This provides us with the quantification of the general criterion stated above. In particular, when selecting the time step, collision detection will be the driving criterion when $\Delta t_d<\Delta t_k$, i.e. when
			\begin{equation}
			\label{eq:v_omega_time}
				v>\frac{4R\varepsilon}{\frac{\pi^2}{4}+\varepsilon^2}\omega_k
			\end{equation}
			and vice versa. Eq.~\eqref{eq:v_omega_time} gives us a useful condition to evaluate the tightest constraint through a simple comparison between maximum velocity and highest frequency in the system. Also, it correlates them with the accuracy of collision detection through $\varepsilon$. As expected, the problem of missing grazing collisions is an issue for high velocities and vice versa. More in detail, the break even point of the criterion is driven by the behavior $v\sim\varepsilon\omega_k$, where $\varepsilon$ is very small and $\omega_k$ is typically very high. Note that the dependency on $R$ shown in~\eqref{eq:v_omega_time} might be misleading, since $R$ appears at the denominator of $\omega_k$ as well. In particular, the mass is proportional to $R^3$ and then the right hand side of Eq.~\eqref{eq:v_omega_time} is proportional to $\sim 1/\sqrt{R}$.
			
			The last point to discuss is about the choice of the contact method (SMC vs NSC vs NSC with compliance) to be used. As mentioned in section~\ref{sub:contact}, SMC are naturally suited for smooth problems (low velocity contacts), NSC for non-smooth problems (high-velocity impacts) while in principle, the NSC method with compliance can be used for both. A more detailed comparison between the aforementioned methods, including simulation of different scenarios, evaluation of performance and criteria for selection of parameters is reported in C::E technical documentation~\citep{Chrono}. Studies to assess the accuracy of such methods are also reported. Unlike gravity, where the exact dynamics of the system are known and can be computed, contact methods do not model in a comprehensive fashion the physics involved in collisional processes. While the results of gravity algorithms can be validated using an analytical exact formulation, the only way to assess the accuracy of contact methods is through experimental test. As mentioned in Section~\ref{sec:validation}, the documentation of C::E~\citep{Chrono} reports the results of several benchmark scenarios, including cone penetration test, standard triaxial test and direct shear test. Further analytical and experimental validation scenarios are reported in~\cite{HeynPhD}.

		\subsubsection*{The gravity-collision problem}
			The full gravitational aggregation problem features both gravitational and collisional dynamics and, in principle, both of them constraint the choice of the simulation time step. However, gravity and collisions act on very different time scales and the characteristic time of gravity is typically several orders of magnitude higher than that of collisions. Also, their dynamics are not coupled. For these reasons, it is not required to perform the simulation using only one time step (which would be the smallest one, i.e. the one related to collisions $\Delta t_c=\min(\Delta t_k,\Delta t_d)$). Instead, as done in~\cite{Sanchez2012}, the aggregation scenario can be simulated using two time steps: the gravity time step $\Delta t_g$, determined using~\eqref{eq:dt_g} and the collision time step $\Delta t_c$, determined using either~\eqref{eq:grazing_coll} or~\eqref{eq:dt_k}. In particular, we advance the simulation of $\Delta t_c$ to properly solve collisional dynamics, but we evaluate gravity (through $N^2$ interactions or by building the octree) every $\Delta t_g$. This allows for consistent savings in terms of computational time, without jeopardizing the accuracy of results.
			
			In this context, a more accurate modeling of the local gravity interactions between fragments would not be beneficial towards a more realistic representation of the granular N-body problem. Past astrodynamics studies have clearly shown that the mutual gravitational dynamics between objects of irregular shape, represented using shape-based (e.g. polyhedron) models, are very different from the case of point-mass sources interaction~\citep[e.g.][]{Fahnestock2006,AIM_FerrariASR}. However, the granular N-body problem is very different from that of astrodynamics applications, which consider purely gravitational motion with no contact and collision interactions. In our case, due to the presence of both gravity and contact interactions, acting at different time scales, a noticeable effect of deviations due to enhanced gravity modeling would appear after a time when many contact/collision interactions have already occurred. In this case, the chaotic nature of the contact/collision interactions represents a much higher source of uncertainty compared to the accuracy of local gravity models. Because of this, all authors identify the better modeling of contact/collision interactions as the way to enhance the realism of simulations \citep[e.g.][and others]{Michel5, Richardson2, Movshovitz2012} and agree to using octrees, at the cost of a lower accuracy of local gravity computations. Unlike astrodynamics applications, the goal here is not to accurately reproduce the coupled and orbital-attitude dynamics of each fragments, but rather to reproduce the global behavior of the system. Relevant parameters concerning gravity are indeed global ones, such as bulk density and total mass of the system, and lower-fidelity local models of gravity (as in octrees) are widely used.
			
	\subsection{Features of the problem}
	\label{sub:features}		
		As discussed, the study of the dynamics of self-gravitating aggregates is a very complex problem and deals with challenges arising from the coupled and chaotic interactions occurring within the granular media. The study of asteroid shapes and rubble-pile dynamics has its fundamentals in the continuum theory, extended by~\cite{Holsapple2001,Holsapple2004,Holsapple2007,Holsapple2010} to the study of rubble-pile objects. However, self-gravitating granular systems are not strengthless fluid bodies and, for example, can spin faster than a perfect fluid before shedding mass~\citep{Richardson2005}. Even for cohesionless aggregates, many effects can contribute to providing strength to the aggregate. These include surface interaction phenomena, such as friction, and other effects related to the geometry of contact interactions between particles such as interlocking~\citep{Sanchez2012}. To summarize all these effects, the strength of a rubble pile can be quantified based on its effective angle of friction, which results from particle shapes, size distributions, packing and surface friction~\citep{Sanchez2012}. \cite{Holsapple2010} showed that the angle of friction is very important to establish the domain of admissible equilibrium shapes and that spins at limiting conditions become larger for a higher angle of friction, implying a larger strength of the aggregate. Accordingly,~\cite{Sanchez2016} observe that the angle of friction inhibits deformation. In~\cite{Zhang2017} geometric interlocking is the main source of shear strength for crystal structures and higher values can be achieved for random packing and increased size heterogeneity of fragments. Among geometrical effects, extreme relevance is given by the angularity of non-spherical particles. As mentioned, spheres behaves very differently from angular particles, for which the contact dynamic problem is very different. Due to their non-smooth shape, angular particles can foster interlocking and between particles. In particular, spheres have only one contact point per pair, while non-spherical objects can have several. This has a great impact on all aspects related to contact dynamics and to the exchange of interactions/forces at contact points. For example, this affects how the value of parameters such as surface friction and restitution coefficients act on a global scale, since they both act at contact points. This means that their effect can be greatly amplified depending on relative geometry. In practice friction and restitution act on a larger scale and contact interactions results more dissipative with respect to the case of spheres. This is confirmed by experiments showing that restitution appears to be much lower for irregular objects~\citep{Hartmann1978} due to the loss of center-of-mass kinetic energy to rotational energy post-collision~\citep{Korycansky2006}. In practice, angular shapes can increase the strength~\citep{Richardson2005,Jiang2015,Zhang2018} of the aggregate and, through a lower restitution coefficient, increase the likelihood of the formation of a stable aggregate during a gravitational re-accumulation process~\citep{Walsh2008}. The resolution of the model, in terms of number of bodies in the rubble pile aggregate, is also very relevant. \cite{Richardson2005} show that coarse configurations consisting of a small number of larger particles are more resistant to tidal disruption or reshaping than fine configurations with many smaller particles, observing that the strength of the aggregate depends on the number of particles as well. A major role in the granular dynamics processes is played by friction. \cite{Sanchez2012} found that friction increases substantially the limiting spin rate. For the case of angular bodies, where the overall effect of friction is higher (many contact points) the expected limiting spin rate would be even higher. \cite{Sanchez2012} also observe that the addition of surface friction makes the aggregate more stable. They conclude that friction might modify the square root dependency between critical spin of a granular aggregate and bulk density. The same is found by~\cite{Zhang2017}, who conclude that higher interparticle friction can keep a spinning rubble pile stable at a higher spin rate. Furthermore,~\cite{Pravec2007} show how maximum spin rate increases (substantially) with friction. To summarize, the chaotic dependency between the dynamical history, initial shape/configuration of the particles, contact/surface parameters and properties of the final aggregate~\citep{Zhang2017} gives a clear picture of the complexity of the problem and the unpredictability of its dynamics.

	\subsection{Simulation examples}
		A set of simulations is performed here to extend results presented by~\cite{Fabio} to an increased number of bodies and to verify the correlations between parameters of the simulation, initial conditions of the N bodies and aggregation outcome. To simulate realistic aggregation scenarios, it is important to carefully select the physical properties of the N bodies and their initial dynamics. Initial conditions play a crucial role to the formation of the aggregates and their properties. The initial dynamical state includes the position and velocity $v_0$ of the center of mass of each body, as well as the angular position and spin rate $\omega_0$ of each body. Typical asteroid aggregation scenarios consider these bodies as fragments created from a parent larger body after, for example, a collision event~\citep{Michel2,CampoBagatin2018}. In such case, the fragments could share a common residual angular momentum due to spinning motion of the parent body. Parameter $\Omega$ is introduced to simulate such effect, and it is defined as the initial residual angular velocity of the system of particles: when $\Omega\ne0$ the fragments have a nonzero orbital velocity around the center of mass of the system. In order to compare and extend the results in~\cite{Fabio}, similar scenarios are set up, using non-smooth dynamics and a time step of 10~s. In our simulations, bodies are convex hulls whose shape is created as the geometrical envelope of randomly generated points. In particular, we use 16 points and corresponding convex hulls have, on average, 10 vertices. Surface friction is based on a simple Coulomb model with coefficient of 0.6. To foster the formation of aggregates, restitution is set to zero (perfectly inelastic collisions). Results presented here refer to simulations with 10$^4$ bodies, with average characteristic size of each fragment of 3~km and maximum initial distance between bodies in the order of 100~km. The physical properties of fragments are chosen among typical values of objects belonging to the main asteroid belt or Near Earth Asteroids population. The material density is set to 3~g/cm$^3$, a common choice for asteroid aggregation simulations~\citep{Sanchez2011}, suitable to produce values of bulk density between 1.2 and 2.5~g/cm$^3$, typical of the asteroid population~\citep{Richardson}. At initial time, the angular and center of mass position of bodies are randomly generated with uniform spatial distribution, as well as the directions of their linear and angular velocity vectors.
		
		\begin{figure}
			\begin{center}
				\includegraphics[width=.325\columnwidth]{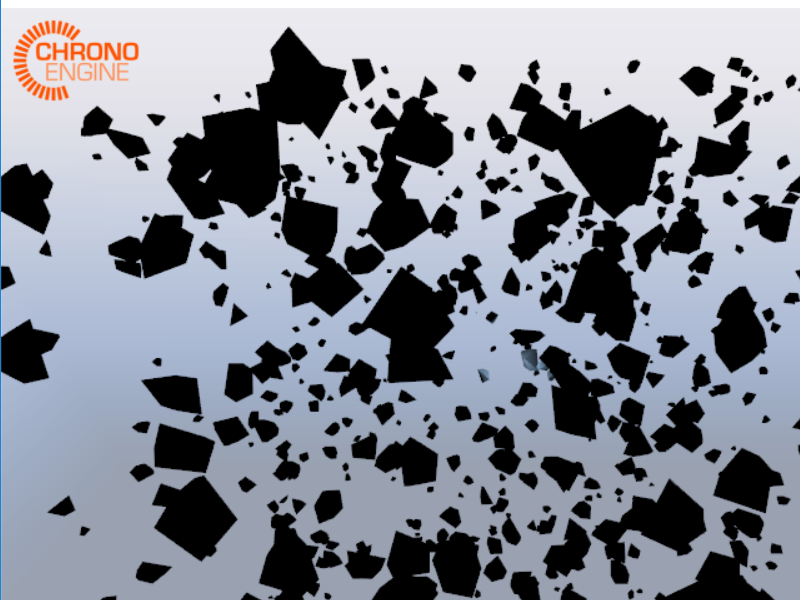}
				\includegraphics[width=.325\columnwidth]{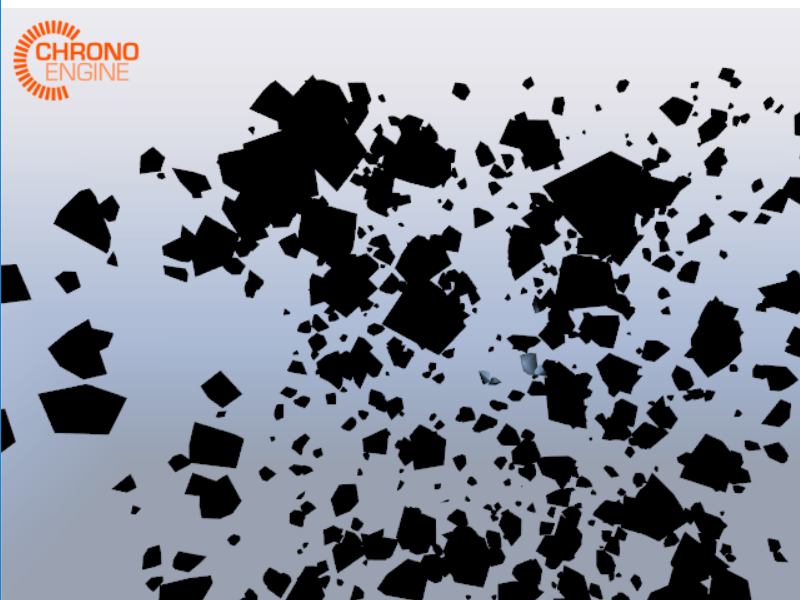}
				\includegraphics[width=.325\columnwidth]{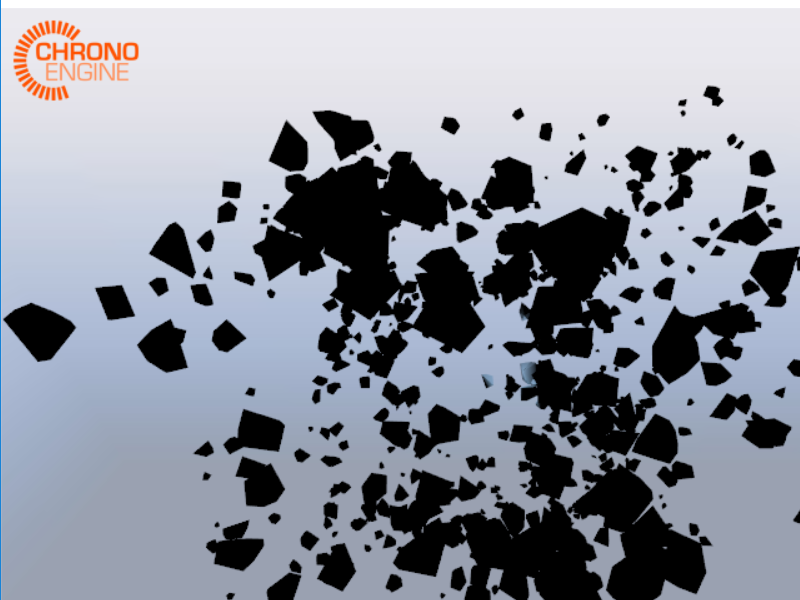}
				\includegraphics[width=.325\columnwidth]{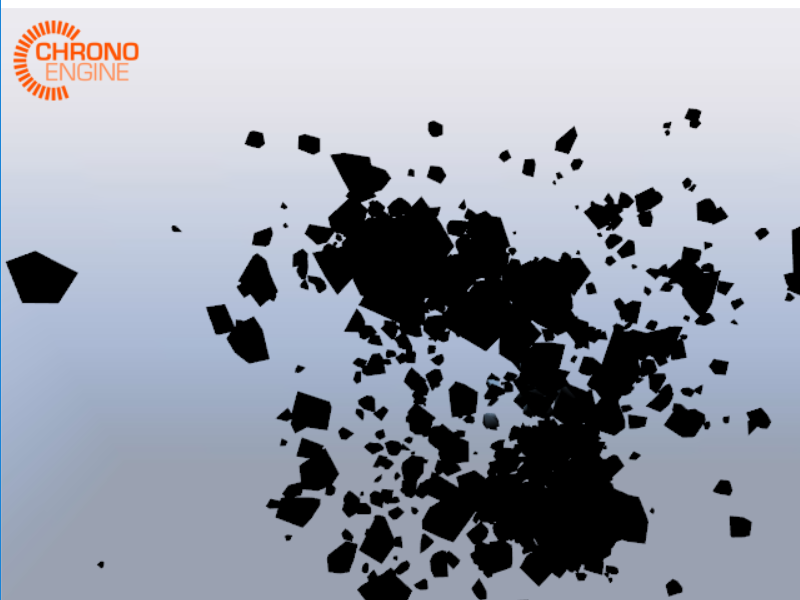}
				\includegraphics[width=.325\columnwidth]{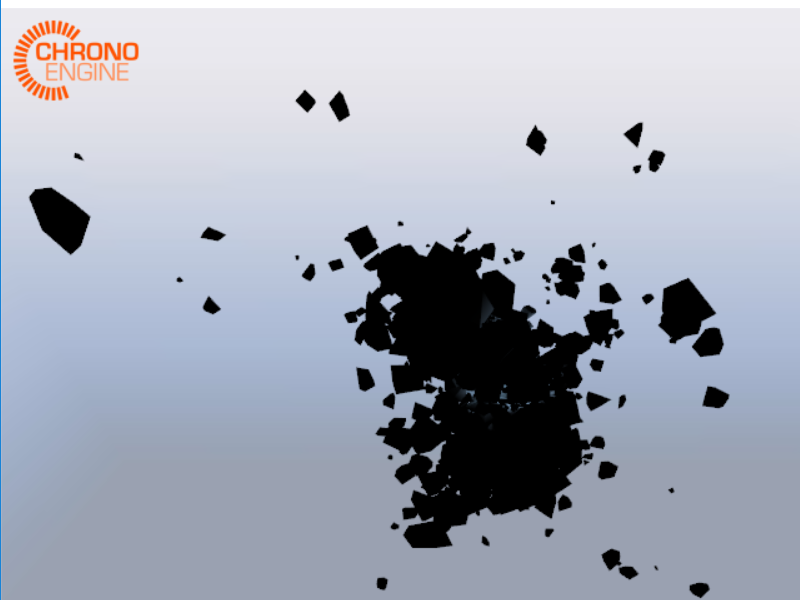}
				\includegraphics[width=.325\columnwidth]{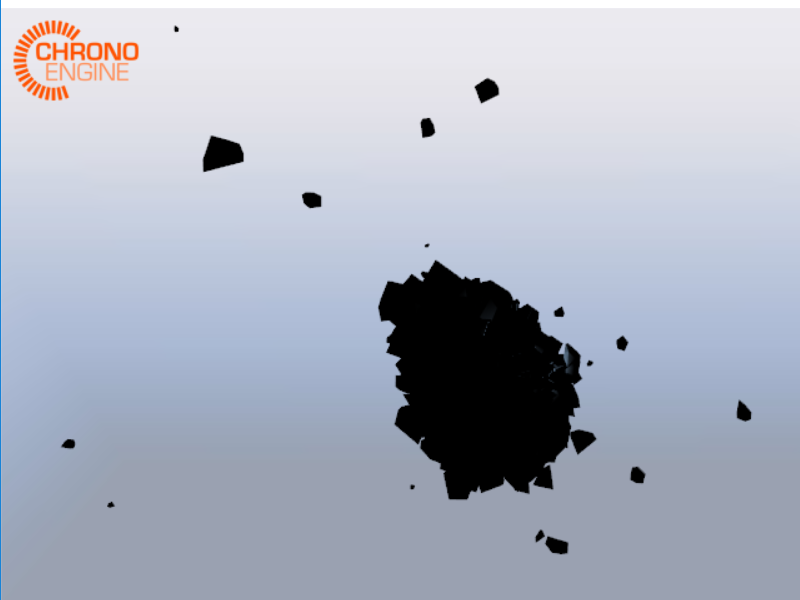}
				\caption[Aggregation sequence example with 1,000 bodies]{Aggregation sequence example with 1,000 bodies}
				\label{fig:aggregation-sequence}
			\end{center}
		\end{figure}

		An example of aggregation sequence is shown in Figure \ref{fig:aggregation-sequence}, for a simplified case with 10$^3$ bodies (to facilitate the visualization of images). While not in contact, the motion of the $N$ bodies is driven solely by their mutual attraction. When fragments start to interact at closer range, few small bodies are scattered away due to collision mechanisms. Depending on initial conditions, one, multiple or no stable aggregates are eventually formed. Aggregation is not observed when relative velocities between bodies are too high and in particular when parameters $v_0$, $\omega_0$ and $\Omega$ are above certain threshold values, above which no aggregation is possible. As for the case under study, these values are identified approximately with $v_0=95$ m/s, $\omega_0=10^{-1}$ rad/s and $\Omega=5\cdot10^{-4}$ rad/s. After reaching a steady state, aggregates are considered as single asteroids. Different kind of aggregates have resulted from the simulation performed.
	
\begin{figure}
			\centering
			\subfigure[\label{fig:aggr-plot1}]
			{\includegraphics[width=0.496\columnwidth]{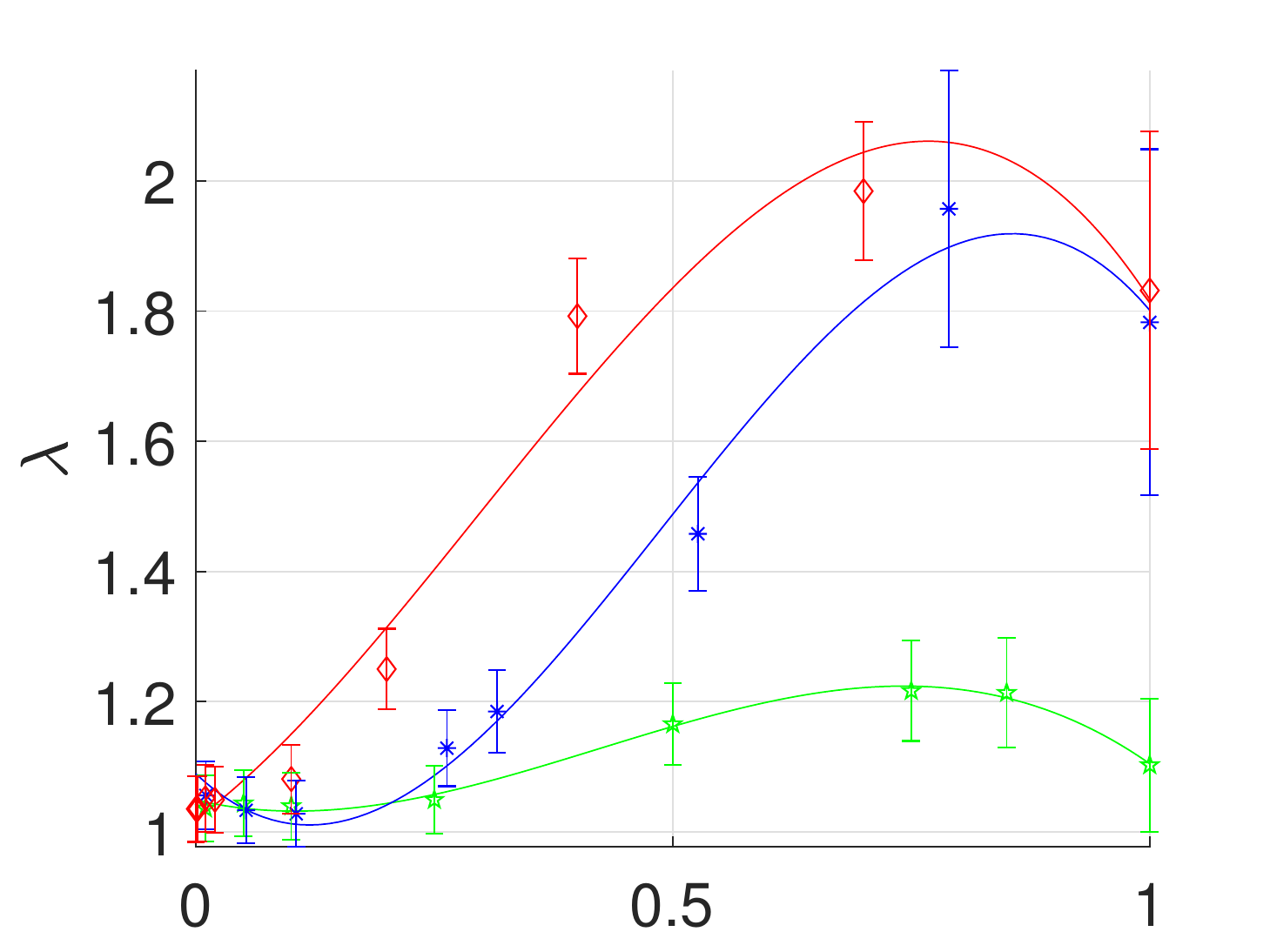}}
			\subfigure[\label{fig:aggr-plot2}]
			{\includegraphics[width=0.496\columnwidth]{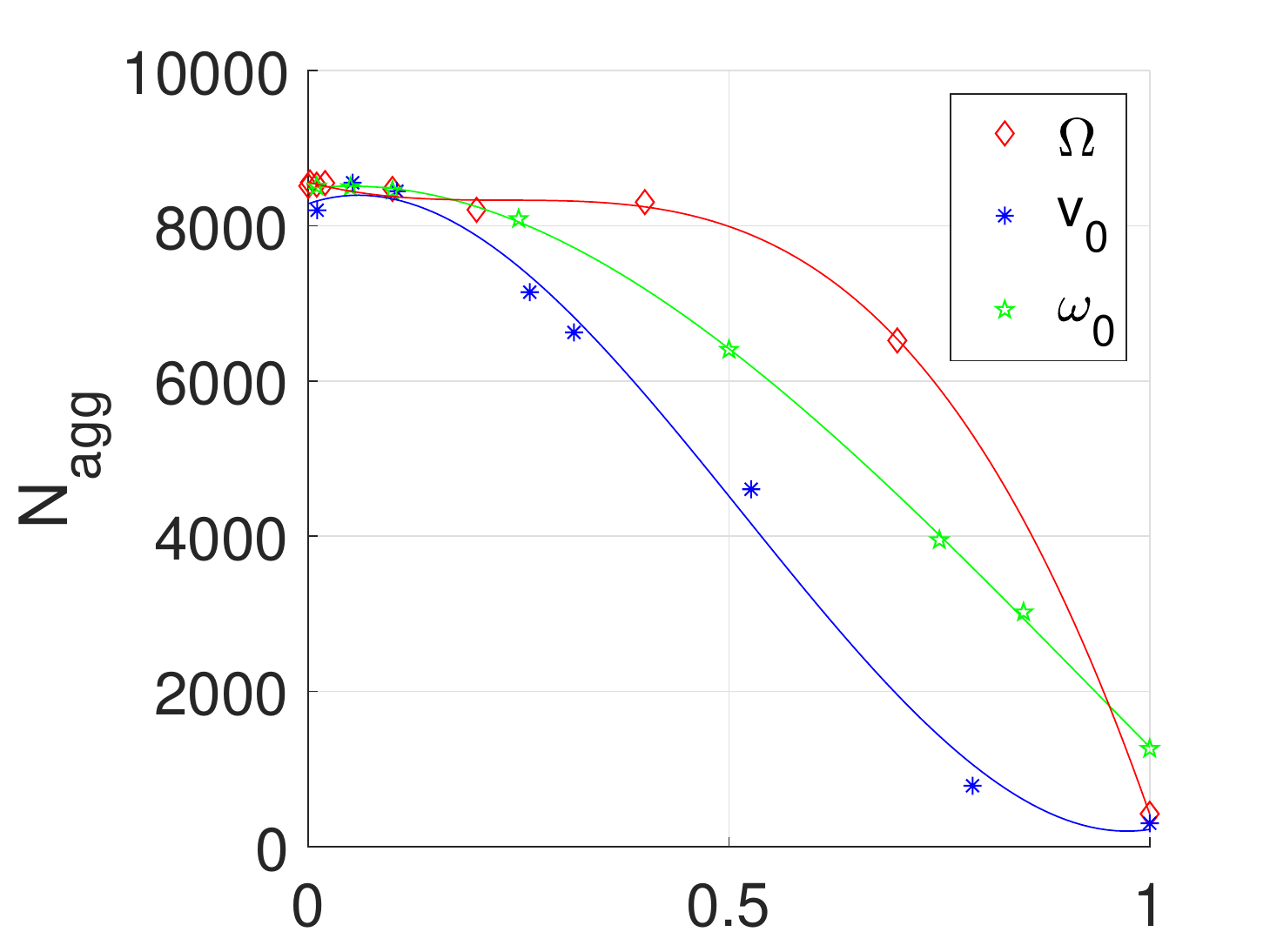}}
			\subfigure[\label{fig:aggr-plot3}]
			{\includegraphics[width=0.496\columnwidth]{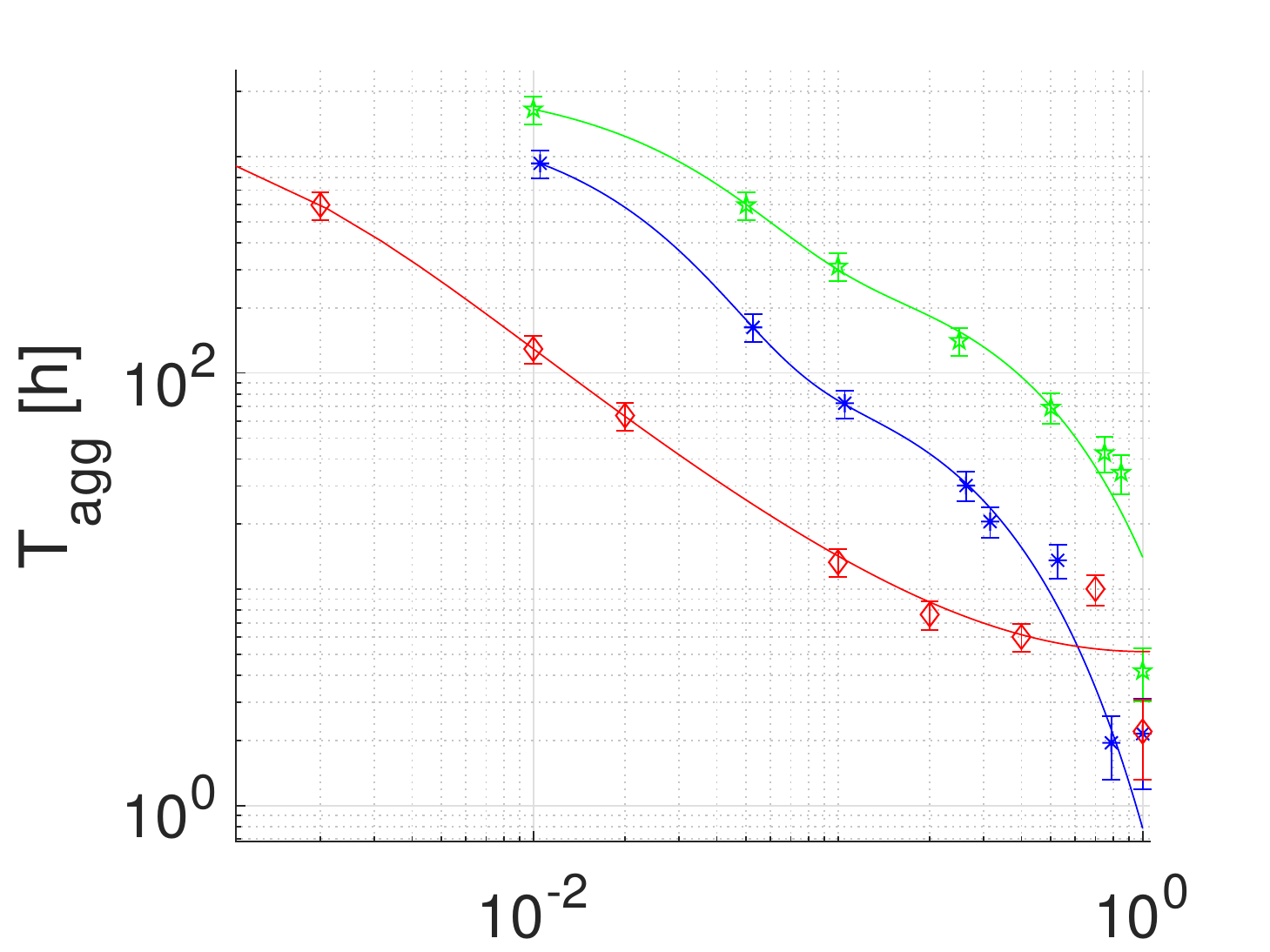}}
			\caption{Properties of the final aggregate as functions of normalized initial conditions: (a) inertial elongation, (b) number of fragments in the main aggregate and (c) rotation period of the main aggregate. Uncertainties due to definition of final aggregate's surface are highlighted.}
			\label{fig:aggr-plot}
		\end{figure}

		Figure~\ref{fig:aggr-plot} shows the properties of the largest aggregate formed after the numerical simulation, as function of the initial dynamics of the system. In particular it shows (a) the inertial elongation $\lambda$, defined as the ratio between maximum and minimum moment of inertia of the aggregate~\citep{Fabio}, (b) the number of fragments in the aggregate $N_{\text{agg}}$ and (c) its rotation period $T_{\text{agg}}$, as function of initial conditions $\Omega$ (red diamonds), $v_0$ (blue asterisks) and $\omega_0$ (green stars). The values of $\Omega,v_0,\omega_0$ are shown on the abscissa normalized to their maximum values, so that each parameter ranges from 0 to 1. Polynomial (for $\lambda$ and $N_{\text{agg}}$) and exponential (for $T_{\text{agg}}$) fitting functions are used to highlight the trend of data distribution.
		
\begin{figure}
			\centering
			\subfigure[\label{fig:aggr-plot4}]
			{\includegraphics[width=0.496\columnwidth]{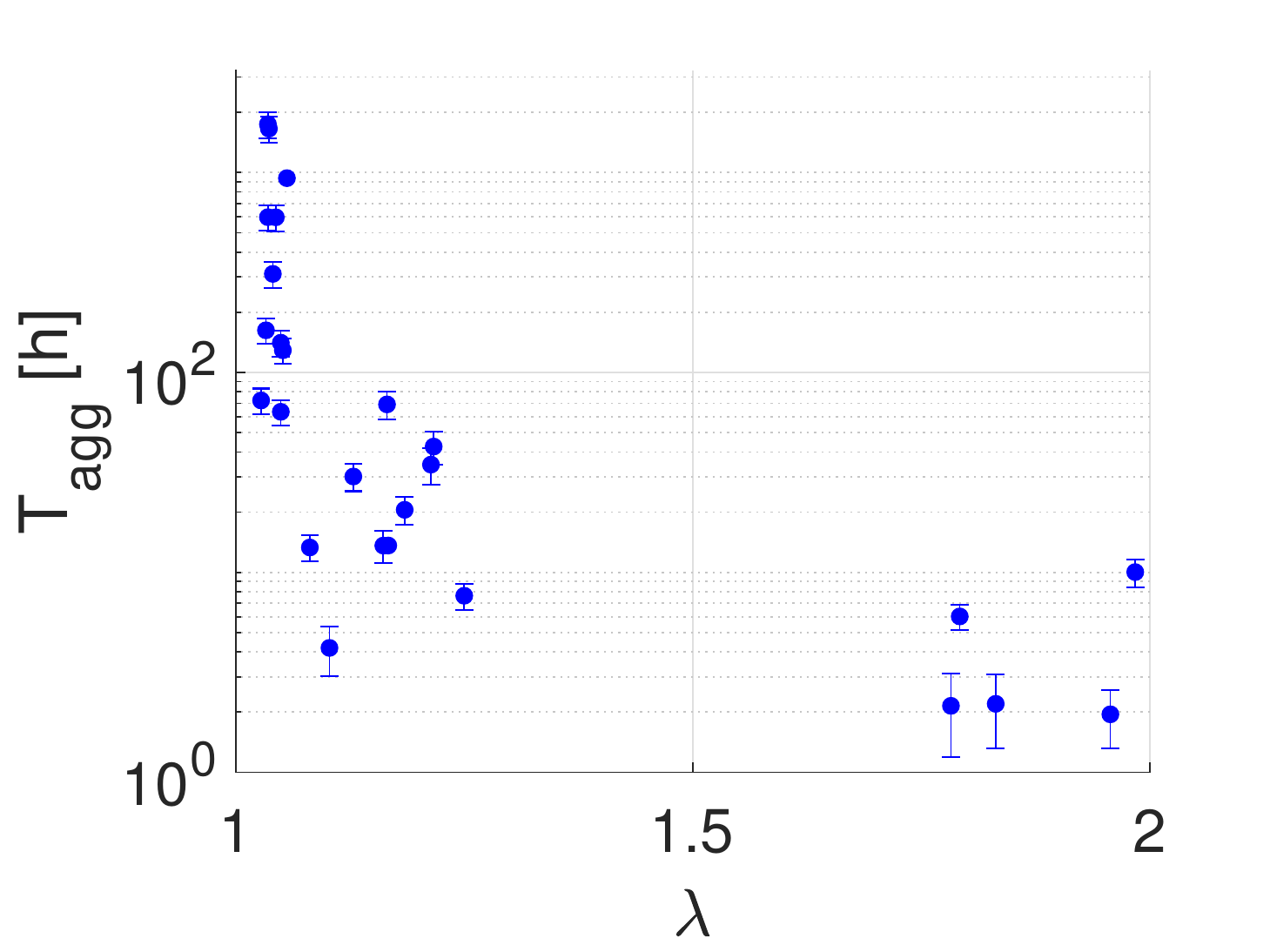}}
			\subfigure[\label{fig:aggr-plot5}]
			{\includegraphics[width=0.496\columnwidth]{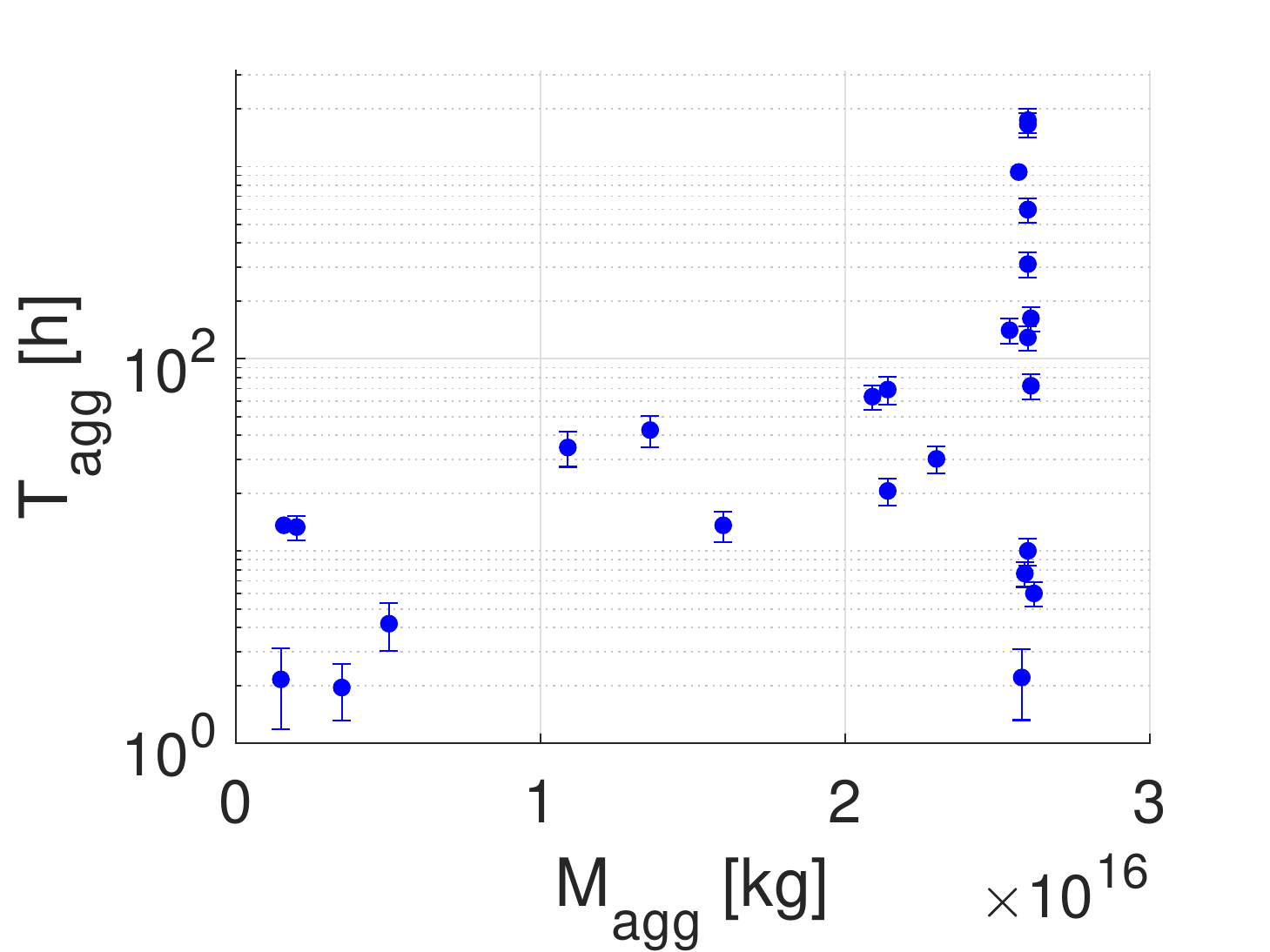}}
			\caption{Rotation period of the final aggregate as functions of (a) inertial elongation, (b) mass of the aggregate. Uncertainties due to definition of final aggregate's surface are highlighted.}
			\label{fig:aggr-plot_T}
		\end{figure}

		The measure of the global properties and geometry of the final aggregate is affected by the uncertainty on the computation of its enveloping surface. As discussed in section~\ref{sub:shape}, quantities such as volume, porosity, bulk density and size of the aggregate are better defined within a range of values computed between a minimum and a maximum volume surface. As a consequence, the inertial elongation $\lambda$ and the angular velocity (or equivalently the rotation period $T_{\text{agg}}$) of the aggregate are affected by the uncertainties of this measurement and should also be defined in within a range of values. In particular, the angular velocity of the aggregate $\omega_{\text{agg}}$ is measured after computing its total angular momentum $H_{\text{agg}}$ and inertia matrix $I_{\text{agg}}$, as $\omega_{\text{agg}}=I_{\text{agg}}^{-1} H_{\text{agg}}$, as in~\cite{Richardson2005}. The computation of the inertia tensor of the aggregate (and its principal inertia axes) rely on the definition of the enveloping surface. The uncertainty of this measurement depends on the size of the single particle with respect to the total aggregate. Hence, the accuracy depends on the number of particles in the aggregate and it is higher for smaller aggregates (those that spin faster in our simulations). \cite{Richardson2005} estimate that the error in axis measurement could be as large as on particle radius along each axis, with errors of about 10\% for the case of 1,000 bodies. When propagated to the computation of angular velocity, they estimate errors larger than 20\%. Similarly, we estimated the error in the computed values of $\lambda$ and $T_{\text{agg}}$, based on the number of bodies in the final aggregate, and show their uncertainties around the computed mean value in Figures~\ref{fig:aggr-plot} and~\ref{fig:aggr-plot_T}. In our case the smaller aggregates are the fastest spinning ones: they are made of tens or few hundreds bodies, meaning potentially very large measurement errors. In other words, the resolution of our smaller aggregates is too coarse and does not allow to estimate accurately their spin rate, elongation, porosity and bulk density. Also, due to their low number of bodies, it is rather inaccurate to use the term rubble pile when referring to such small aggregates. To a closer inspection, the structure of these aggregates is closer to a monolith rather than a rubble pile: they are made of one (or few) big boulders and tens of smaller particles resting on their surface. This structure is by no means a typical self-gravitating aggregate and it's rather closer to the ``test particle on a rigid sphere'' case reported in~\cite{Weidenschilling1981} and~\cite{RichardsonQuinn}, where critical spin may be estimated using the simple balance between gravitational and centrifugal force~\citep{Hestroffer2019}. This is indeed the result we find for them. The spin rate for aggregates with many more particles are comparable to what has been found by other authors within the error prescribed by the uncertainties mentioned above. Among these, our fastest spinning aggregates have spin rates in the order of 4.2$\pm$0.9 h.

		General trends of Figures~\ref{fig:aggr-plot} and~\ref{fig:aggr-plot_T} agree with expected behavior and are coherent with what discussed in section~\ref{sub:features}. As expected, higher elongations are obtained for higher values of $v_{0}$, $\omega_{0}$ and $\Omega$. Maximum elongation is reached for values between 75\%-85\% of their limiting value. For higher values of initial conditions ($>$75\%-85\%) the simulations result in the formation of multiple aggregates of smaller size, which are more typically regularly shaped than larger ones. It is indeed observed that maxima in Figure~\ref{fig:aggr-plot1} represent thresholds after which the formation of more than one significant aggregate occurs. As confirmed in Figure~\ref{fig:aggr-plot2}, the higher the velocities, the lower number of bodies is found in the main aggregate, which is smaller for higher values of $v_{0}$, $\omega_{0}$ and $\Omega$. Figure~\ref{fig:aggr-plot3} confirms a well-known result, with larger aggregates having a higher rotation period (lower spin rate) with respect to smaller ones.
		Figure~\ref{fig:aggr-plot_T} shows the rotation period of the final aggregate as function of (a) inertial elongation and (b) mass of the aggregate. As expected, fast rotators are more elongated and less massive, while slow rotators have a more regular rounded shape and are more massive.

\section{Conclusion}
	The paper presents the numerical implementation of a code for coupled gravity-granular dynamics problems and reports numerical methods available to solve gravitational and collisional dynamics. We build upon the work by~\cite{Fabio} and extend the capabilities of the code by introducing a parallel CUDA-GPU octree structure, to be able to evaluate mutual gravity for a higher number of bodies. Unlike existing N-body codes, the implementation presented here has the capability to handle non-spherically shaped bodies to a full double precision accuracy. Compared to classical sphere-based codes, this allows for a more realistic simulation of contacts between bodies and open to new opportunities and scenarios to be simulated. After presenting the results of test scenarios to validate newly implemented modules, we discuss the features and performance of numerical methods and derive requirements arising from the dynamics of asteroid aggregation scenarios. The problem of numerical simulation of gravitational aggregation is addressed: we provide guidelines and quantify criteria to properly set up such numerical simulations. We show examples of possible applications by performing numerical simulations of gravitational aggregation. The parameter space explored includes initial conditions of bodies in terms of relative velocity between them, spinning rate and residual angular velocity of the system. The outcome of the analysis is discussed showing the properties of the final self-gravitating aggregate after its stable formation in terms of inertial elongation, period of rotation and number of bodies in the aggregate.

\section*{Acknowledgements}
	This project has received funding from the European Union's Horizon 2020 research and innovation programme under the Marie Skłodowska-Curie grant agreement No 800060. Part of the research work was carried out at the Jet Propulsion Laboratory, California Institute of Technology, under contract with the National Aeronautics and Space Administration. The authors would like to thank the anonymous reviewers for their comments and suggestions that helped to increase the quality of the paper.
	



\bibliographystyle{mnras}
\bibliography{mybibfile} 




%
%


\bsp	
\label{lastpage}
\end{document}